\let\latexaddtocontents\addtocontents
\documentclass[twocolumn, a4paper, final, accepted=2023-07-04]{quantumarticle}
\let\addtocontents\latexaddtocontents
\pdfoutput=1

\usepackage{booktabs}

\usepackage{tikz}
\usepackage{graphicx}
\usepackage{amsmath, amsfonts}
\usepackage{hyperref}
\hypersetup{
    citecolor=cyan,
    colorlinks=true,
    linkcolor=cyan,
    urlcolor=cyan,
    }
\usepackage{braket}

\usepackage{xcolor}
\usepackage{listings}

\definecolor{codegreen}{rgb}{0,0.6,0}
\definecolor{codegray}{rgb}{0.5,0.5,0.5}
\definecolor{codepurple}{rgb}{0.58,0,0.82}
\definecolor{backcolour}{rgb}{0.95,0.95,0.92}

\lstdefinestyle{mystyle}{
    backgroundcolor=\color{backcolour},
    commentstyle=\color{codegreen},
    keywordstyle=\color{magenta},
    numberstyle=\tiny\color{codegray},
    stringstyle=\color{codepurple},
    basicstyle=\ttfamily\small,
    breaklines=true,
    numbers=left,
}
\usepackage{amsthm}
\usepackage[capitalise]{cleveref}

\usepackage{authblk}
\usepackage[sorting=none, style=numeric-comp]{biblatex}
\usepackage{caption}
\usepackage{subcaption}

\usepackage{algorithmicx}
\usepackage{algorithm}
\usepackage{algpseudocode}
\usepackage{titlecaps}
\usepackage{mathtools}

\def\Tr{\mbox{tr}}
\def\hc{^{\dagger}}

\def\>{\rangle}
\def\<{\langle}

\def\E{ {\mathcal E} }

\def\U {{\mathcal U}}

\def\N{ {\mathcal N} }

\def\D{ {\mathcal D} }
\def\T{ {\mathcal T} }
\def\I{ \mathbbm{1} }

\newtheorem{theorem}{Theorem}[section]

\newtheorem{lemma}[theorem]{Lemma}
\newtheorem{definition}[theorem]{Definition}

\newcommand{\random}{Random Circuits}
\newcommand{\pauli}{Pauli-Gadget Circuits}

\newcommand{\supremacy}{quantum computational supremacy}
\newcommand{\X}{\textsf{X}}
\newcommand{\SX}{\textsf{SX}}
\newcommand{\Z}{\textsf{Z}}
\newcommand{\Y}{\textsf{Y}}
\renewcommand{\T}{\textsf{T}}

\newcommand{\RZ}{\textsf{RZ}}

\DeclarePairedDelimiter{\floor}{\lfloor}{\rfloor}

\newcommand{\CX}{\textsf{CX}}

\newcommand{\curlbrac}[1]{\left\{ #1 \right\}}
\newcommand{\sqrbrac}[1]{\left[ #1 \right]}
\newcommand{\brac}[1]{\left( #1 \right)}

\newcommand{\inout}[2]{
	\vspace{7pt}
	\hspace*{\algorithmicindent} \textbf{Input:} #1 \\
	\hspace*{\algorithmicindent} \textbf{Output:} #2

	\hrulefill
	}

\usepackage{tcolorbox}
\usepackage{tabularx}
\usepackage{array}
\usepackage{colortbl}
\tcbuselibrary{skins}

\newcolumntype{Y}{>{\raggedleft\arraybackslash}X}
\newcolumntype{S}{>{\hsize=.5\hsize}X}
\newcolumntype{O}{>{\hsize=.7\hsize}X}
\newcolumntype{P}{>{\hsize=1.5\hsize}X}

\tcbset{tab2/.style={ width = 18cm, enhanced,fonttitle=\bfseries,fontupper=\normalsize\sffamily,
		colback=yellow!10!white,colframe=red!50!black,colbacktitle=blue!40!white,
		coltitle=black,center title}}

\usepackage{csquotes}

\title{Volumetric Benchmarking of Error Mitigation with Qermit}
\author[1,4]{Cristina Cirstoiu}
\orcid{0000-0002-7341-5261}
\author[1,4]{Silas Dilkes}
\orcid{0000-0003-2186-0379}
\author[1,4]{Daniel Mills} 
\orcid{0000-0001-5902-3774}
\author[1]{Seyon Sivarajah}
\orcid{0000-0002-7332-5485}
\author[1,2,3]{Ross Duncan}
\orcid{0000-0001-6758-1573}
\affil[1]{Quantinuum, Terrington House, 13-15 Hills Road, Cambridge CB2 1NL, UK}
\affil[2]{Department of Computer and Information Sciences, University of Strathclyde, 26 Richmond Street, Glasgow G1 1XH, UK}
\affil[3]{Department of Physics and Astronomy, University College London, Gower Street, London, WC1E 6BT, UK}
\affil[4]{These authors contributed equally:\{cristina.cirstoiu, silas.dilkes, daniel.mills\}@quantinuum.com}

\begin{document}

\sloppy

\maketitle

\begin{abstract}
    The detrimental effect of noise accumulates as quantum computers grow in
    size. In the case where devices are too small or noisy to perform error correction, error
    mitigation may be used. Error mitigation does not increase the fidelity of
    quantum states, but instead aims to reduce the approximation error in
    quantities of concern, such as expectation values of observables. However,
    it is as yet unclear which circuit types, and devices of which
    characteristics, benefit most from the use of error mitigation. Here we
    develop a methodology to assess the performance of quantum error
    mitigation techniques. Our benchmarks are volumetric in design, and are
    performed on different superconducting hardware devices. Extensive
    classical simulations are also used for comparison. We use these benchmarks
    to identify disconnects between the predicted and practical performance of
    error mitigation protocols, and to identify the situations in which their
    use is beneficial. To perform these experiments, and for the benefit of the
    wider community, we introduce \emph{Qermit} -- an open source python package for
    quantum error mitigation. Qermit supports a wide range of error mitigation
    methods, is easily extensible and has a modular graph-based software
    design that facilitates composition of error mitigation protocols and
    subroutines. 
\end{abstract}

\section{Introduction}

Noise inhibits the development of quantum processors \cite{preskill2018quantum,
harrow2017supremacy}. Techniques for reducing the level and effects of noise
have been proposed, and act at each layer in the quantum computing stack. At
the hardware level, noise reduction may be achieved through calibration
\cite{sheldon2016procedure}, dynamical decoupling \cite{dynamicl1999viola},
pulse-level optimisation \cite{carvalho2021error}, etc.  At the highest level
of abstraction, fault-tolerant methods for encoding and processing information
with logical qubits may be used \cite{shor1995scheme, terhal2015error}. 

Between these two extremes there exists two classes of noise reduction
techniques. The first are those that increase process fidelity through
operations at compile time. This class includes noise-aware routing and qubit
allocation \cite{murali2019noiseadaptive, Sivarajah_2020}, noise tailoring
\cite{wallman2016noise}, and circuit optimisation \cite{Sivarajah_2020}.  The
second are those which do not increase the fidelity of prepared states
directly, but instead improve accuracy when measuring quantities of concern,
such as expected values of observables.  Several such \emph{error mitigation}
protocols have been proposed, including Zero-Noise Extrapolation (ZNE)
\cite{temme2017error, li2017efficient, giurgica2020digital}, Probabilistic
Error Cancellation (PEC) \cite{temme2017error}, Clifford Data Regression (CDR)
\cite{Czarnik2021errormitigation}, Virtual Distillation
\cite{huggins2021virtual}, and many others \cite{endo2020hybrid,
bharti2022nisqreview,koczor2021exponential}. 

Importantly, both of these classes of noise reduction techniques require less
information about the experimental setup than do hardware level approaches.
Additionally, neither compile time error reduction nor error mitigation
protocols require a significant overhead in the number of qubits, as in the
case of error correction \cite{fowler2012surface, ogorman2017magic}. This often
comes at the cost of an increase in the number of circuits and shots in the
case of error mitigation. 

The importance of noise suppression at the hardware and logical levels is clear
as both are vital to the development of scalable fault-tolerant quantum
processors. Compile time approaches to noise reduction have also been
extensively benchmarked \cite{mills2020applicationmotivated} and shown to
perform strongly. What’s unclear is the impact of error mitigation. While proof
of principle experiments have shown improved accuracy when using these
techniques \cite{temme2017error, li2017efficient, Czarnik2021errormitigation,
giurgica2020digital}, there has been no systematic study of their practical
effectiveness and limitations.  As such it is unclear which circuit dimensions
and types, nor quantum computing devices of which characteristics, are well
suited to the use of error mitigation. Indeed, the differing assumptions made
about the underlying noise by each error mitigation protocol may manifest as
unpredictable practical performance. 

We clarify the practical utility of error mitigation by introducing and
implementing a methodology for the benchmarking of error mitigation protocols.
Our benchmarking experiments compare the accuracy of operator expectation value
calculations when a selection of error mitigation protocols are employed, and
when they are not. 

The design of our benchmarks is inspired by volumetric benchmarks of quantum
processors \cite{BlumeKohout2020volumetricframework}, allowing us to estimate
the ‘volume’ of the circuit depth and width where the error mitigation
protocols perform well.  The particular circuits used are constructed from a
variety of application-motivated circuit classes
\cite{mills2020applicationmotivated}. This ensures our benchmarks are
indicative of practical performance, which we use to identify classes of
computations where error mitigation may be used fruitfully.  Taking a
volumetric approach ensures that results of these benchmarks may be quickly
compared, and that a wide range of circuit sizes are covered. In this work, we
benchmark CDR and ZNE, but our method is applicable to a broad class of error
mitigation protocols.

We conduct our experiment on real quantum computing devices, and using
classical simulators. Each error mitigation method makes different assumptions
about the underlying noise, and as such it is difficult to compare error
mitigation methods using noisy simulations alone without introducing bias. This
is particularly true since existing noise models are poor predictors of
hardware behaviour beyond a few qubits. Indeed our experiments on real hardware
demonstrate reduced performance as compared with noisy classical simulation,
across different error mitigation methods. 

For the benefit of the community, and in order to conduct these benchmarking
experiments, we have developed \emph{Qermit}. Qermit is an open-source python
package for the design and automated execution of error mitigation
protocols. The error mitigation
protocols presently available through Qermit include those explored in our
benchmarking experiments, namely ZNE and CDR. Qermit also includes
implementations of PEC, error mitigation based on frame randomisation
\cite{wallman2016noise}, and protocols performing correction through
characterisation of State Preparation And Measurement (SPAM) errors. In all
cases, several variations of each protocol are provided. Qermit provides a
common interface to this selection of error mitigation schemes, simplifying
their use. Further, Qermit supports the straightforward construction and
combination of error mitigation protocols and sub routines, facilitating quick
prototyping of new protocols. By virtue of being implemented using Tket
\cite{Sivarajah_2020}, Qermit is platform-agnostic, and so may be used with a
wide range of quantum hardware, and in conjunction with several common quantum
software development kits.

The remainder of this paper is structured as follows. In \cref{sec:em} we give
an overview of error mitigation and the particular schemes that we investigate
in this work. In \cref{sec:qermit} we introduce Qermit and details of the
implementations of CDR and ZNE. In \cref{sec:benchmark} we introduce the design
of our benchmarks, and the philosophy that motivates them.  In
\cref{sec:results} we give the results of the experiments we have conducted.
Finally we conclude in \cref{sec:conclusion}.
 	
\section{Quantum Error Mitigation }
\label{sec:em}

Error mitigation protocols typically have many steps in common. In \cref{sec:em
framework} we describe a general framework for error mitigation
of observables which takes into account the practical aspects of their
implementation on quantum hardware. Recent works \cite{takagi2021fundamental,
cai2021practical} have also exploited these common features to analyse the
overhead in sample complexity, and to determine (universal) lower bounds. Our
approach focusses on the modularity of error mitigation protocols, also
exploited by the design of Qermit, and takes into account all quantum and
classical resources required for an error mitigation experiment. In
\cref{sec:em zne} and \cref{sec:em cdr} we use this framework to describe ZNE
and CDR. In \cref{sec:em noise} we describe the noise profile assumptions that
justify the design choices made when developing ZNE and CDR.

\subsection{Unifying Framework}
\label{sec:em framework}

Consider a target input quantum circuit given by a sequence of gates $U = U_1 U_2
...U_d$, an initial (pure) state $\rho_0$, and an observable $O$.\footnote{This
	may also be extended to a class of circuits or a set of observables.} The error
mitigation protocols studied here output an estimator $\<\hat{O}\>_{EM}$ of the
true expectation value $\<O\>= \Tr( U\rho_0 U\hc O)$. This estimator should
reduce the noise bias in the estimation of this quantity on the backend. 

To achieve this, error mitigation protocols run the given circuit and/or other
quantum circuits on \emph{backends}, such as quantum processing units or
classical simulators, which may be noisy or ideal. The outputs from backends
are binary strings, called \emph{shots}, which may be combined to produce, for
example, expectation values. 

In generating the estimator $\< \hat{O} \>_{EM}$, many error mitigation
protocols employ the following steps.

  	\begin{description}
  		\item[Data Collection:] 
  		The first step consists of a noise characterisation procedure $\mathfrak{N}$ that takes as input the target experiment(s) $ (U,\rho_0, O)$, along with a set of resource parameters $\mathcal{R}$. $\mathcal{R}$ can include: the total number of distinct circuits $K$, shots per each circuit $(n_i)_{i=1}^{K}$, and allowed qubits; the type and amount of classical simulation used; and the backend $q$ with fixed specifications such as the compilation strategy, architecture, and noise features.
  		
  		$\mathfrak{N}$ involves (i) a series of sub-processes $\mathfrak{N}_{1}$, ..., $\mathfrak{N}_{K}$, each of which modify the input circuit $U$ in a method-specific way 
 		and (ii) measurement circuits $\mathfrak{M}_1(O), ...,  \mathfrak{M}_M(O)$ with classical estimator function $o_{m,i}(z)$ for $m=1,...,M$ and $i=1,...,k$ where $z$ labels the measurements outcomes.
 
  		The data collection step returns a set of (labelled) complex parameters given by $ {\bf{D}} = ({D}^{q}_{i,m}, {D}^{c}_{i,m})$, indexed by each of the sub-processes, where 
  		
  		\begin{equation}
  			D_{i,m}^{q} =   \frac{1}{n_i} \sum_{s=1}^{n_i} \sum_{z} o_{m,i}(z) Z_s(z)
  			\label{eqn:estimator}
  		\end{equation}	  			 		  	
  		with $Z_s$ an independent identically distributed (i.i.d) indicator random variable over measurement outcomes obtained from evaluating $\mathfrak{M}_m(O)\mathfrak{N}_i(U)(\rho_0)$ on the quantum device $q$. If required and available, $D_{i,m}^c$ corresponds to the exact classical simulation. 
  		\item[Functional Model:] An implicit mapping $\mathfrak{F}$ (or a set thereof) between the output parameters of the noise characterisation step and the (unknown) error mitigated estimator $\<\hat{O}\>_{EM}$ so that
  		\begin{equation}
  			\mathfrak{F}({\bf{D}}_{}, \<\hat{O}\>_{EM}) = 0. 
  		\end{equation}
  		The specific form of this function is typically motivated by assumptions on the noise characteristics of the quantum device.  
  		\item[Data Processing:] This step, is completely classical and aims to produce an
  		output estimator $\<\hat{O}\>_{EM}$ based on fitting the data $\bf{D}$ to the functional model $\mathfrak{F}$. Depending on the particular function this may be a simple summation or a classical optimisation algorithm to determine the coefficients of $\mathfrak{F}$.
  	\end{description}

    All the processes involved in producing $\mathbf{D}$ may be described in
    the quantum combs formalism, which generalises quantum channels to higher
    order operations \cite{chiribella2008quantum}.  
    $\mathfrak{N}$ will generally depend on the type of noise a particular
    method is aiming to mitigate. For example, a set of the sub-processes and
    measurements may be independent of $U$ or $O$, with the aim to produce
    (partial) tomographic information \cite{temme2017error}.
    The framework also allows for adaptive processes, which is
    to say that $\mathfrak{N}_i$ may depend on outcomes $D_{1,m}$...
    $D_{{i-1},m}$ for a subset of the measurement circuits $m$. Typically the
    Data Collection step will include the identity process, which does not
    modify the circuit $U$ or observable, and produces a noisy (sample mean)
    estimator $\<\hat{O}\>_{N}$ of the expectation value of the observable
    given resources $\mathcal{R}$.

Note that several factors at each step can influence the performance of an
error mitigation technique.  An example is the accuracy in the noise modelling
assumptions that motivate the choice of functional model, and we discuss this
further in the case of ZNE and CDR in \cref{sec:em noise}. The accuracy of the
data collected, influenced by the available shot budget, also impacts the
accuracy and variance in the final error mitigated observable expectation
approximation. We explore this variance in the case of our experiments in
\cref{ap:finite-sampling-fit}.

 	 \subsection{Zero Noise Extrapolation}
     \label{sec:em zne}
 	 The Zero Noise Extrapolation (ZNE) method \cite{temme2017error, li2017efficient} assumes that noise may be controlled by a parameter $\lambda$ (or more generally a set of parameters) which can be viewed as a proxy for average gate error rates. Then, for a given quantum circuit, the noisy expectation value of a target observable will be a function depending on $\lambda$. One may produce different samples of this function by artificially increasing the noise parameter to different levels $\lambda_1, \lambda_2,...,\lambda_k$. An extrapolation process then produces an estimate of the expected value for $\lambda=0$, the ideal case where no physical errors occur. 
 	 
 	 There are several ways in which one may boost physical errors affecting a quantum circuit. One method involves increasing the duration of pulses involved in producing each gate within the circuit \cite{temme2017error}. A second approach, which we review here, is to introduce additional gates to obtain a higher depth but equivalent unitary circuit \cite{giurgica2020digital}. 
 	 
 	 \paragraph{Data Collection:} ZNE involves a series of sub-processes  $\mathfrak{N}_{\lambda_1}, ..., \mathfrak{N}_{\lambda_k}$, that take the input circuit $U=U_1 U_2 \, ... \, U_d$ and produce a modified, or \emph{folded} \cite{giurgica2020digital}, circuit 
 	 \begin{equation}
 	 	\mathfrak{N}_{\lambda_i}(U) = U_1 (C_1^{} C_1\hc)^{\alpha_1^{i}} U_2 (C_2^{} C_2\hc)^{\alpha_2^{i}} \, ... \, U_d ( C_d^{} C_d\hc)^{\alpha_d^{i}}.
 	 \end{equation}
 	 Here  $C_i C_i\hc = \mathbb{I}$ so as to not introduce logical error, and $\alpha^{i}_j$ are positive integers. There is flexibility in the choice of the $C_i$ unitaries, and \cite{giurgica2020digital} analyse different folding variations: (i) circuit folding when only $\alpha_d^{i} \neq 0$ and $C_d = U_1 ... U_d$, (ii) random gate folding with $C_j=  U_j$ and $\alpha_{j}^{i}$ chosen uniformly at random for a fixed noise level $\lambda_i = \sum_{j}(2\alpha_j^{i} + 1)$.  
 	 
 	 The output of the first step will be $\mathbf{D} := (D_{\lambda_1}, D_{\lambda_2}, ..., D_{\lambda_k})$ where each entry is a sample mean estimator for the expectation value $\Tr( O \mathfrak{N}_{\lambda_i}(U)\rho_0 [\mathfrak{N}_{\lambda_i}(U)]\hc )$ of the target observable with respect to the modified circuit at each noise level,  given the fixed set of resources (i.e number of shots).
 	
 	\paragraph{Functional Model:} In the case of ZNE, there are several possibilities for the data-fitting functional. We outline several below which have been explored in Refs.~\cite{temme2017error, giurgica2020digital}.
 	
 	a) The \emph{polynomial extrapolation} assumes that the dependency on the noise level parameter $\lambda$ can be expressed as a truncated Taylor series so that the data is fitted to the function
 	\begin{equation}
 	D_{\lambda} = 	\<\hat{O}\>_{EM}  +  \sum_{i=1}^{K} F_i \lambda^{i},
 	\label{eqn:poly}	
 	\end{equation}	
	with negligible higher order terms $O(\lambda^{K+1})$ and unknown (complex) parameters $F_i$. If the number of data points, or equivalently in our case the number of noise parameters $k$, is at least $K+1$, the number of unknown parameters, then the extrapolation is well defined. In the special case when $k=K+1$ there exist analytic expressions for the coefficients, and the method is referred to as Richardson extrapolation \cite{endo2020hybrid}.
	
 	 b) The \emph{exponential extrapolation} assumes the expected values of observables decay exponentially with the noise level parameter $\lambda$ and the data can then be fitted to
 	 \begin{equation}
 	 	D_{\lambda} = \<\hat{O}\>_{EM} + F (e^{-f\lambda} - 1). 
 	 	\label{eqn:exp}
  	 \end{equation}
   
      c)  The \emph{poly-exponential extrapolation} assumes that the exponential decay with the noise level has a polynomial expansion so it is fitted to the function
      \begin{equation}
      	D_{\lambda} = \<\hat{O}\>_{EM}  + F(e^{-\sum_{i=1}^{K} F_i \lambda^i }-1)
      	\label{eqn:polyexp}
      \end{equation}

 	 \subsection{Clifford Data Regression}
 	 \label{sec:em cdr}
 	 Clifford Data Regression (CDR) \cite{Czarnik2021errormitigation} is a machine learning approach to error mitigation. The method relies on the idea that circuits containing a number of \T{} gates logarithmic in the number of qubits can be efficiently simulated \cite{bravyi2016improved}. 
 	 
 	 \paragraph{Data Collection:} CDR involves a series of sub-processes $\mathfrak{N}_0(U) = U$ and $ \mathfrak{N}_1,..., \mathfrak{N}_k$, each of which modifies the input $U$, synthesized into a universal Clifford + T gate set, to produce a circuit where all except a small number $N_{nc}$ of $\T{}$ gates are replaced by a single-qubit Clifford gate  $\{\I{}, \rm{S}, \rm{S}\hc, \Z{}\}$. The resulting unitary circuits $\mathfrak{N}_i(U)$ are efficiently simulated classically and so measurements of $O$ will involve both quantum and classical evaluation. The output will be $\mathbf{D} = ( D_0^{q}, (D_{1}^{q}, D_{1}^{c}), ..., (D_{k}^{q}, \tilde{D}_{k}^{c}))$ where each pair consists of the ideal classically simulated expectation value of the target observable $O$ with respect to the modified circuit  $D_{i}^{c} = \Tr( O \mathfrak{N}_{i}(U)\rho_0 [\mathfrak{N}_{_i}(U)]\hc )$ and respectively $D_{i}^{q}$ a corresponding sample mean estimator evaluated on quantum device with resources $\mathcal{R}$.
 	 
 	 \paragraph{Functional Model:} The functional that relates an error mitigated estimate $\<\hat{O}\>_{EM}$ of the ideal expected value $\<O\> = \Tr(OU\rho_0 U\hc)$ to the data obtained in the noise characterisation is 
 	 \begin{equation}
 	 	\<\hat{O}\>_{EM} = f( D_{0}^{q}),
 	 \end{equation}
 	 where $f = {\rm{min}}_{g\in \mathcal{F}} || \mathbf{D}^{c} - g(\mathbf{D}^{q}) ||_{2}^{2}$   minimises the distance measure   
 	 $||\mathbf{D}^c - g(\mathbf{D}^{q})||_2^{2} =  \sum_{i=1}^{k} [ D_i^c - g(D_i^q)]^2$ over all (invertible) functions $g$ in a specified class $\mathcal{F}$.
 	 
 	 In particular, in \cite{Czarnik2021errormitigation} the class of fitting functions $\mathcal{F}$ was assumed to be linear so that $f(x) : = F_1 x+ F_0$ and (assuming $F_1\neq 0$)
 	  \begin{equation}
 	 	\<\hat{O}\>_{EM} = F_1 D_0^{q} + F_0.
 	 	\label{eqn: CDRmodel}
 	 	\end{equation}
 
 	 \subsection{Noise Profile Assumptions }
     \label{sec:em noise}
 	 The use of different fitting functions in ZNE and CDR are motivated by an incoherent, Markovian noise model. Such a model is described by the quantum channel
 	 $\mathcal{N} = (1-\lambda) \mathcal{I}  + \lambda \mathcal{E}$
 	 with $\mathcal{I}$ the identity operation, and $\mathcal{E}$ an arbitrary process.
 	 Therefore, the noisy implementation of the target unitary channel $\mathcal{U}(\cdot) := U (\cdot ) U\hc$ is given by $\tilde{\U}:=\U_1\circ \N\circ \U_2\circ \N ...\circ \U_d \circ \N$.  We can expand this out in terms of linear combinations of channels with coefficients depending on the noise parameter  $\lambda$ to get $\tilde{\U} =  (1-\lambda)^{d}\,\U + \lambda(1-\lambda)^{d-1} \binom{d}{1}\E^{(1)} + .... + \lambda^{d}\, \mathcal{E}^{(d)}$, where the notation $\E^{(k)}$ is an average over all processes in the expansion of $\tilde{\U}$ that have exactly $k$ errors $\E$. 

     In total there are $\binom{d}{k}$ different ways $k$ errors occur within the circuit. Therefore, $\tilde{\U}$ and the corresponding noisy expectation value can be expressed in terms of a power series in $\lambda$ with degree at most $d$, thus motivating the polynomial fit. Alternatively, the analysis in \cite{cai2021multiexponential} considers approximating the binomial coefficients in the expansion of $\U$ with a Poisson distribution so that $ \lambda^{k} (1-\lambda)^{d-k} \binom{d}{k} \approx e^{-d\lambda} \frac{(d\lambda)^k}{k!}$ and therefore,  this noise model with the assumption that $\lambda = O(1/d)$  makes the approximation  $\tilde{\U} \approx e^{-d\lambda}[\U+ \sum_{k=1}^{d} \frac{(d\lambda)^k}{k!} \bar{\E}^{(k)} ]$ valid. The noisy expectation value will then take a similar form  which motivates the use of exponential and poly-exponential fitting functions. 
 	 In particular, for a global depolarising noise model (where $\mathcal{E}$ outputs a fixed state $\psi_0$) the noisy and exact expected values have a linear relationship 
 	 \begin{equation}
 	 \<O\>_{N} = (1-\lambda)^{d} \<O\> + (1- (1-\lambda)^d)\Tr(O\psi_0).	
 	 \end{equation}
  	In such a case, the linear fit in CDR and polynomial fit in ZNE (with $K\leq d$) are not susceptible to noise bias so the true expectation can be recovered exactly, up to finite sampling size errors.


 	

\section{Qermit Implementation Details}
\label{sec:qermit}

Qermit has a graph based architecture design, which splits the execution of
error mitigation methods into atomic functional processes. These sub-processes
may be the submission of a circuit to a QPU, the modification of a circuit for
the purposes of error mitigation, or the fitting of models. A graph, with these
processes as vertices, is specified to describe how the processes should
exchange inputs and outputs.

Qermit is complementary to Mitiq \cite{larose2021mitiq}, which is an
open-source Python toolkit that implements an overlapping set of error
mitigation schemes. Qermit takes a different approach that breaks-down the
implementation of each protocol into standalone modular units, with a graph
describing how each module interacts defined separately. This allows the
modules to be easily re-used, modified and composed.

This software architecture allows for vertices to be amended to adapt the
protocol where necessary, and relieves developers of some of the work of piping
together processes. A modular design additionally means sub-graphs and graphs
may be reused in other original error mitigation protocols, which is
particularly useful as error mitigation schemes typically have several steps in
common. As a result it is ensured that Qermit: is easily extensible; allows
one to readily combine error mitigation protocols; and facilitates quick
prototyping of new protocols through the reuse of existing sub-protocols.
Combining error mitigation protocols has been shown to be a fruitful endeavour
\cite{bultrini2021unifying, mari2021extending, cai2021multiexponential}, and
one which is simplified in Qermit.

\subsection{\texttt{MitRes} and \texttt{MitEx}}

There are two types of error mitigation methods in Qermit: \texttt{MitRes}
methods, which modify the distribution of shots retrieved from a backend; and
\texttt{MitEx} methods, which return a modified expectation value estimator of some
observable. In general a \texttt{MitRes} or \texttt{MitEx} object may perform
any modification matching this form, or no modification at all. The instances
of \texttt{MitRes} and \texttt{MitEx} objects in Qermit are designed to perform
error mitigation.

\texttt{MitRes} and \texttt{MitEx} objects are constructed as dataflow graphs,
called a \texttt{TaskGraph}. Each node of a \texttt{TaskGraph} is a
\texttt{MitTask} object; itself a function that computes some step or
sub-process of an error mitigation protocol. Edges of the \texttt{TaskGraph} move
data between \texttt{MitTask} objects which depend on each other. When the
\texttt{run} function is called on a \texttt{MitRes} or \texttt{MitEx} object,
a topological sort is applied to the graph to order these tasks sequentially.
The tasks are then run locally in this sequential order.

In its default construction, as displayed in \cref{fig:qermit mitres
taskgraph}, a \texttt{MitRes} object will simply run each circuit through a
backend and gather the results. Similarly, the default construction of a
\texttt{MitEx} object, as displayed in \cref{fig:qermit mitex taskgraph}, will
simply estimate the expectation of each desired observable without applying any
mitigation methods.  To do so it will modify the circuit, appropriately
rotating the measurement basis according to the target observable. An example of
such a use of the default \texttt{MitRes} object can be seen in \cref{sec:code
mitres}, and of the default \texttt{MitEx} object in \cref{sec:code mitex
default}.

\begin{figure}
	\centering
	\includegraphics[scale=0.55]{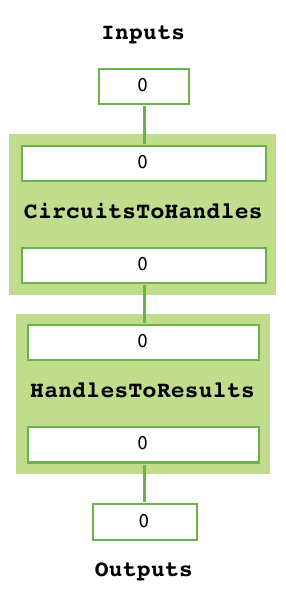}
	\caption{\textbf{\texttt{MitRes} \texttt{TaskGraph}.} The \texttt{CircuitsToHandles}
		\texttt{MitTask} takes circuits and a number of shots as input. It ensures
		the inputted circuits adhere to the requirements of the backend, then
		submits the circuits to the backend; returning identifying results handles.
		\texttt{HandlesToResults} uses the inputted results handles to retrieve
		results from the backend, retuning them as outputs.}
	\label{fig:qermit mitres taskgraph}
\end{figure}

\begin{figure}[t!]
	\centering
	\includegraphics[scale=0.55]{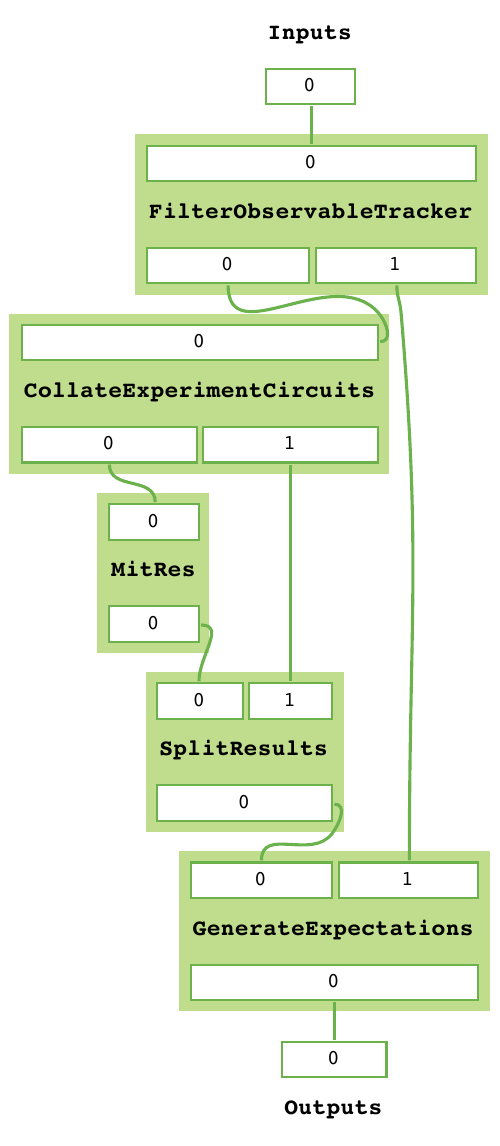}
	\caption{\textbf{\texttt{MitEx} \texttt{TaskGraph}.} The \texttt{MitRes} task may be
		expanded as in \cref{fig:qermit mitres taskgraph}.  Other \texttt{MitTask}
		objects featuring here include: \texttt{FilterObservableTracker}, which
		takes as input a description of the circuit to be implemented, and the
		observable to be measured, returning circuits modified to perform the
		measurements necessary to calculate the expectation of the requested
		observable; \texttt{CollateExperimentCircuits}, which reformats the list of
		circuits to facilitate parallelism; \texttt{SplitResults}, which undoes
		this reformatting; and \texttt{GenerateExpectations}, which uses the results
		to calculate the requested expectation values.}
	\label{fig:qermit mitex taskgraph}
\end{figure}

\subsection{ZNE in Qermit}
\label{sec:qermit zne}

Several options exist in Qermit for both noise scaling and extrapolation to the
zero noise limit. Each noise scaling operation available is of the digital
variety, and includes: circuit folding, random gate folding, and odd gate
folding. Extrapolation functions available include: exponential, linear,
poly-exponential, polynomial, and Richardson. Those chosen for our experiments
are discussed in \cref{sec:results}.

\begin{figure}[t!]
    \centering
    \includegraphics[width=0.9\linewidth]{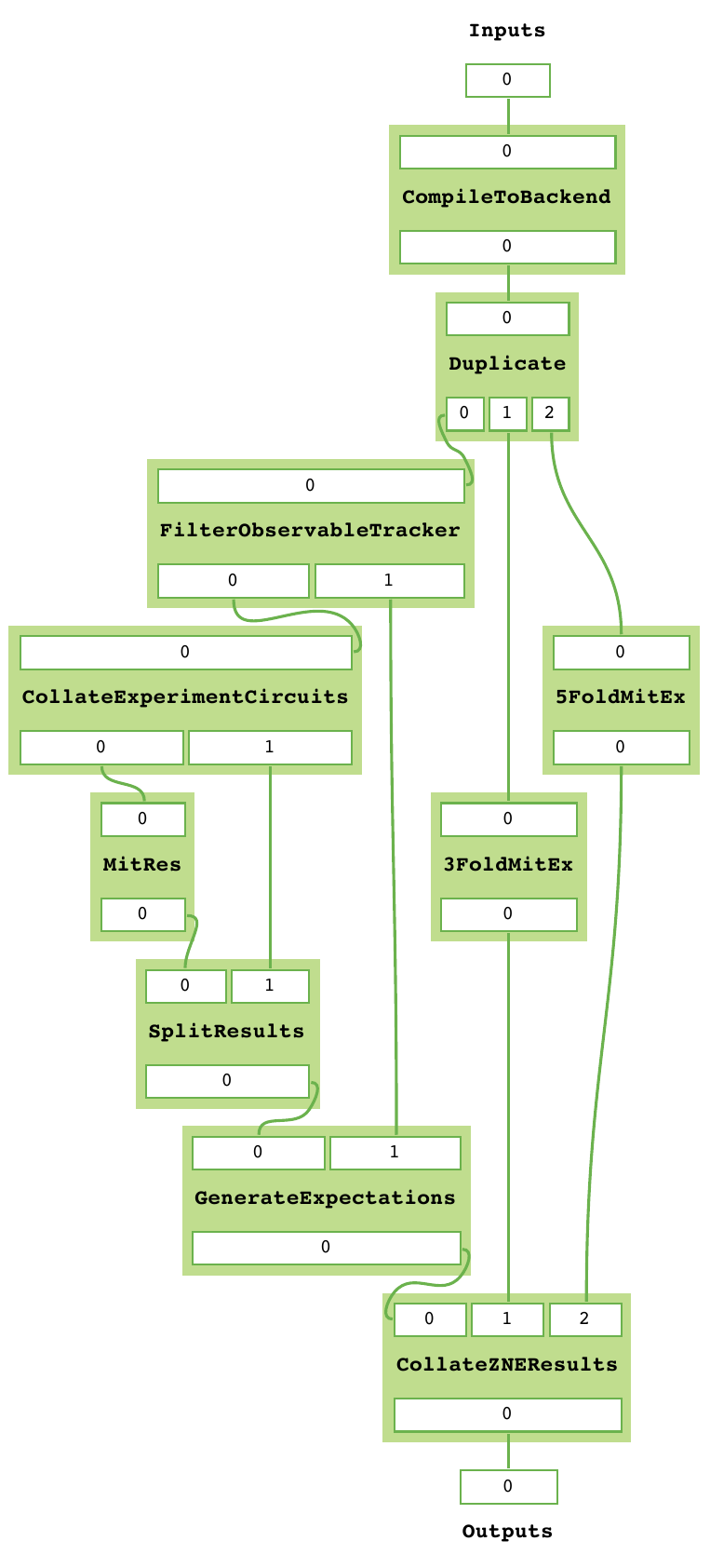}
    \caption{\textbf{ZNE \texttt{TaskGraph}.} This \texttt{TaskGraph} includes:
    \texttt{CompileToBacked}, which ensures the circuit obeys the connectivity
    and gate set restraints of the backend used; \texttt{Duplicate}, which
    created copies of the inputted circuit; \texttt{iFoldMitEx}, which
    increases the noise in the circuit by a factor $i$, and runs the resulting
    circuit using a \texttt{MitEx} of \cref{fig:qermit mitex taskgraph}; and
    \texttt{CollateZNEResults}, which gathers the experiment results and
    extrapolates to the zero noise limit. Note that other \texttt{MitTask}
    objects appearing in this figure have parallels in \cref{fig:qermit
    mitex taskgraph}. Indeed the \texttt{iFoldMitEx} contains a \text{MitEx}
    \texttt{TaskGraph}, which we do not expand for succinctness, as well as a
    \texttt{MitTask} perfoing the noise scaling.}
    \label{fig:qermit zne taskgraph}
\end{figure}

An example \texttt{TaskGraph} which corresponds to a ZNE \texttt{MitEx} can be
seen in \cref{fig:qermit zne taskgraph}. One notices in particular that the
initial circuit is duplicated by the \texttt{Duplicate} \texttt{MitTask},
before each duplicate is passed to a \texttt{iFoldMitEx} \texttt{MitTask}. Each
\texttt{iFoldMitEx} increases the noise in the circuit by a factor of $i$, and
runs the resulting circuit. The original circuit is also passed through a
default \texttt{MitEx}, as displayed in \cref{fig:qermit mitex taskgraph}. The
results are collated and used to produce an error mitigated expectation value
by the \texttt{CollateZNEResults} \texttt{MitTask}. An example of a use of ZNE
within Qermit can be seen in \cref{sec:code mitex zne}.

\subsection{CDR in Qermit}
\label{sec:qermit cdr}

One approach to generating the near Clifford training circuit set required for
CDR uses Markov chain Monte Carlo techniques \cite{Czarnik2021errormitigation}.
Here a first training circuit is produced, where all but a fixed number of
non-Clifford gates are randomly replaced with nearest Clifford gates according
to a weighted distribution. Subsequent steps generate circuits sequentially by
randomly replacing \texttt{$n_{pairs}$} of the Clifford gates in the previously
generated circuit with their original non-Clifford, and similarly
\texttt{$n_{pairs}$} non-Clifford gates with (nearest) Clifford gates. At each
step the new training circuit is accepted/rejected according to a (usually
problem-dependent) user-defined maximal likelihood function. A second option,
implemented in Qermit, and used in the experiments of \cref{sec:results}
replaces uniformly at random \texttt{$n_{pairs}$} of non-Clifford/Clifford
pairs in the first training circuit, without producing a chained training set.

\begin{figure}
    \includegraphics[width=0.9\linewidth]{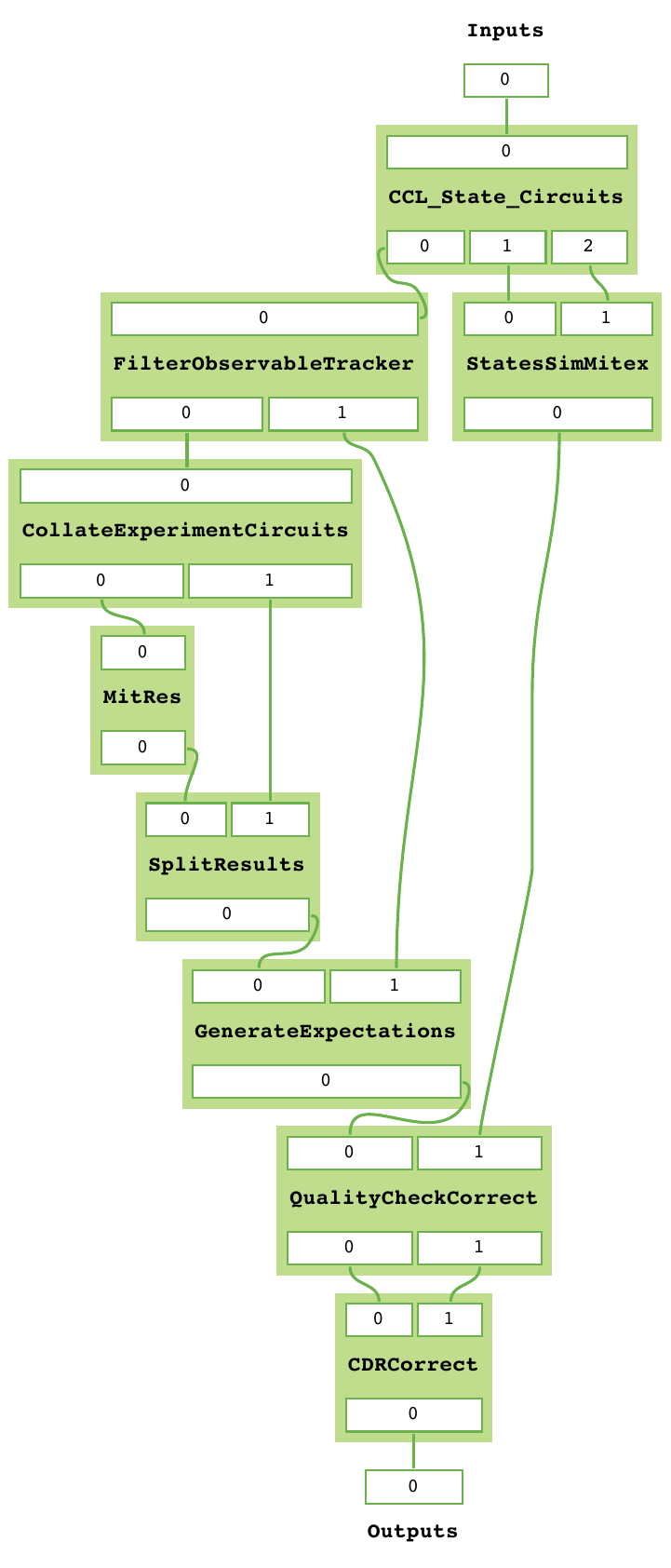}
    \caption{\textbf{CDR \texttt{TaskGraph}.} Besides those \texttt{MitTask}
    objects which are common to \cref{fig:qermit mitex taskgraph}, this
    \texttt{TaskGraph} additionally includes: \texttt{CCLStateCircuits},
    which outputs the original circuits as output 0, and training circuits
    copied as outputs 1 and 2;
    \texttt{StatSimMitEx}, which runs these two sets of circuits through a given
    backend and an ideal classical simulator, returning the results;
    \texttt{QualityCheckCorrect}, which assesses the expected quality of the
    resulting function model; and \texttt{CDRCorrect}, which uses the results from the
    original and training circuits to find a function model and produce an
    error mitigated result. Note that \texttt{StateSimMitEx} would expand to
    reveal a \texttt{MitEx} for the inputted backend, and a \texttt{MitEx} for
    an ideal classical simulator.}
    \label{fig:qermit cdr taskgraph}
\end{figure}

Parameters specifying an instance of a CDR \texttt{MitEx} include the device
backend, the classical simulator backend, the number of non-Clifford gates in
the training circuits, the number of pair replacements, and the total number of
training circuits.  An example \texttt{TaskGraph} which corresponds to a CDR
\texttt{MitEx} can be seen in \cref{fig:qermit cdr taskgraph}.  The initial
circuit is transformed by \texttt{CCLStateCircuits} to prepare it for three
different experiments.  Respectively these are to: run the original circuit on
the given backend, run training circuits with a reduced number of non-Clifford
gates on the same given backend, and run the training circuits on an ideal
classical simulator. Once these experiments have been conducted, a series of
checks conducted in \texttt{QualityCheckCorrect} ensure that the training data gives a
well-conditioned (least-squares) optimisation.  \texttt{CDRCorrect} finds the
fit parameters and produces an error mitigated estimator. An example of the use
of CDR can be found in \cref{sec:code mitex cdr}.

\section{Benchmark Design}
\label{sec:benchmark}

Here we develop a benchmark procedure which can be used to compare error
mitigation methods on different backends.  To
assess the performance of an error mitigation protocol, we introduce the
relative error of mitigation in \cref{sec:benchmark error}. We further discuss in \cref{sec:benchmark performance} factors that influence such performance. Our benchmarks are volumetric by design, with the required circuit structure,
presentation style and results interpretation discussed in \cref{sec:benchmark
volume}. Finally we detail the precise circuits we use in \cref{sec:benchmark
circuits}.
 
\subsection{Accuracy Measures for Error Mitigation}		
\label{sec:benchmark error}
In order to assess the performance of an error mitigation strategy we propose the following criteria that such a performance metric should satisfy (i) faithful -- takes a fixed (zero) value when the error mitigated estimator matches the exact value (ii) operational -- relates to improvement in a computational task  (iii) efficiently estimable from experimental data and classical processing --  allows for scalability.

\begin{definition}	
	Let $O$ be an observable with ideal expectation $\<O\>$ with respect to the
	target state $\psi$, $\<\hat{O}\>_{N}$ a sample mean estimator for the corresponding noisy state $\rho$ and $\<\hat{O}\>_{EM}$ the sample mean estimator for the expectation value after applying the error mitigation method.
    The \emph{absolute error} and \emph{absolute error of mitigation} are
    respectively
    \begin{align}
    	\epsilon_{N} & =  |\<\hat{O}\>_{N} -  \<O\>|\\
    	\epsilon_{EM} & =  |\<\hat{O}\>_{EM} -  \<O\>|
    \end{align}
    
    The \emph{relative error of mitigation} is defined as
    \begin{equation}
    	\epsilon(O)  =  \frac{|\<\hat{O}\>_{EM} -  \<O\>|}{|\<\hat{O}\>_N - \<O\>|}.
    \end{equation}
	\label{def:errors}
\end{definition}

The intuition for the above definition is that we want
a measure that expresses how much closer to the true value is the mitigated
estimator compared to the noisy expectation. 
Note that the relative error of mitigation has the following operational
properties (i) faithful $\epsilon = 0$ iff corrected expectation values
matches the exact value and (ii) whenever $\epsilon \leq 1 $ the mitigation
of errors was successful. 

Generally, the measures in \cref{def:errors} involve a classical simulation computing the true expectation value, and therefore are limited to determine performance of error mitigation schemes in these regimes. In \cref{app:certify} we discuss the problem of certifying error mitigation for target states or circuit classes for which there is no available classical simulation. We introduce several performance metrics that satisfy (i) - (iii) and are based on mirroring, or efficiently simulable circuits. For example, mirrored circuits give scalable benchmarks since the exact expectation values depend only on the input (product) state and observable itself; no classical simulation of the circuits is needed in this case. We make use of this approach to benchmarking in \cref{sec:results}. However, we leave for future work to determine how well these measures predict the general performance of error mitigation on quantum hardware, particularly in regimes approaching quantum advantage.

Furthermore, since both $\<\hat{O}\>_{EM}$ and $\<\hat{O}\>_{N}$ are estimators they incur a variance due to finite sampling statistics. In general the ratio of two random variables does not have a well defined variance and the ratio of two normally distributed variables with zero mean gives rise to the Cauchy distribution, which is typically heavy tailed. However, under mild conditions one can show \cite{diaz2013existence} that $\epsilon$ can be approximated by normal distribution with mean $\mu =\frac{\mu_{EM}}{\mu_N}$ and variance	
\begin{equation}
    \label{equ:relative error variance}
	\sigma_{\epsilon}^2 = \frac{\sigma_{EM}^2}{(\mu_{N}-\<O\>)^2} + \frac{\sigma_N^2 \, \, (\mu_{EM}-\<O\>)^2}{(\mu_N-\<O\>)^4}
\end{equation}
where $\sigma_{EM}^2$ and $\sigma_{N}^2$ are the variance of estimators $\<\hat{O}\>_{EM}$ and $\<\hat{O}\>_N$ due to finite sampling statistics and $\mu_{EM} = \mathbb{E} \<\hat{O}\>_{EM} $, $\mu_{N} = \mathbb{E}\<\hat{O}\>_N $  correspond to the means in the limit of an infinite number of samples. We discuss in \cref{ap:relative-error} the conditions under which the above approximation holds.

\subsection{Performance of Error Mitigation Methods}
\label{sec:benchmark performance}

There are a series of factors that can affect the performance of error mitigation methods. 
Firstly, finite sampling effects are amplified in the error mitigated estimator which incurs a higher variance than the noisy expectation value estimator for a fixed number of shots. This limitation has previously been discussed in Ref.~\cite{endo2020hybrid} and recent work \cite{takagi2021fundamental} derives theoretical lower bounds on the sample complexity overhead of error mitigation in terms of the maximal noise bias and distinguishability measures. In particular, for a local depolarising noise model the number of shots required to produce a mitigated estimator with the same accuracy as the noisy estimator scales exponentially with the circuit depth \cite{takagi2021fundamental, wang2021can}. Practically, if the architecture has a restricted topology then this scaling can even depend exponentially on the number of qubits for sparser graphs that require an $O(n)$ routing overhead \cite{cqcrouting}. 

Secondly, the functional model used to produce the error mitigated estimator will generally not fully capture the underlying backend noise effects. This is particularly restrictive for real hardware, where $\<\hat{O}\>_{EM}$ will therefore be susceptible to noise bias.

Thirdly, fitting parameters to the functional model involves a classical optimisation that may be ill conditioned or unstable, partly due to the increased variance in the finite sampling or in the functional fit. 

Finally there are several specific regimes where the above issues can be more detrimental to the performance of error mitigation strategies. For example if low levels of noise occur then $\<\hat{O}\>_N$ already produces a good estimator of $\<O\>$ with high accuracy and due to the additional sampling overhead error mitigation will not, on average, improve upon that estimator for a fixed shot budget. Typically error mitigation strategies involve reconstructing the surface defined by $\mathfrak{F}$ from the noisy samples $\bf{D}$ in the data collection step -- however, if the finite sample error dominates then solutions to the fit parameters will be unstable. This situation can occur for high levels of noise and is typically exacerbated when the target observables have low expected values. In the following section and \cref{ap:relative-error} we make these observations more precise.

\subsubsection{Amplification of finite sampling variance  in the error mitigated estimator} 

Recall that the error mitigated estimator $\<\hat{O}\>_{EM}$ is produced by classical post-processing which fits the experimental results to a functional model $
\mathfrak{F}({\bf{D}}_{}, \<\hat{O}\>_{EM}) = 0 $. The terms $D_{i,m}^q$ are sample mean estimators that will introduce finite sampling effects. In the simplest case one might have a linear functional where $\<\hat{O}\>_{EM} = \sum_{i,m} F_{i,m} D_{i,m}^{q}$ for some (real) coefficients $F_{i,m}$ and where  $D_{i,m}^{q}$ are defined in \cref{eqn:estimator}. The variance due to finite sampling in the error mitigated estimator is then
\begin{align}
	\sigma_{EM}^{2} :&= Var [\hat{O}_{EM} ] = \sum_{m,i} F_{i,m}^2 Var[D_{i,m}^{q}] \\
	& =  \sum_{m,i} \frac{F_{i,m}^2 o_{m,i} }{n_i} \sigma_{i}^2
\end{align}
where $o_{m,i} := \sum_z o_{m,i}(z)^2$ are constants pre-determined from the target observables (and any modified observables required in the data collection step) and recall that $n_i$ are the number of independent samples $\{Z_{s}\}_{s=1}^{n_i}$ each with variance $\sigma_i^2$.

\subsection{Volumetric Benchmarking}
\label{sec:benchmark volume}


Volumetric benchmarking of error mitigation assesses the overall performance of a method on a specific backend with a fixed set of total resources $\mathcal{R}$. We employ circuit classes with increasing depth $d$ (as determined by the number of layers of primitive circuits) and qubit number $n$ (which is to say the number of qubits the circuit acts on). Our methodology is inspired by volumetric benchmarks of quantum processors \cite{BlumeKohout2020volumetricframework} and consists of:

\begin{enumerate}
	\item Select a class of circuits $\mathcal{C}(n,d)$ and a probability distribution or method to sample $C$ individual circuits. 
	\item Select a (Pauli) observable $O$ (or set thereof) with fixed locality.
	\item Determine relative error of mitigation $\epsilon_{i}(O)$ for each circuit $ \mathcal{C}(n,d)$ labelled by $i\in\{1,..., C\}$.
	\item Determine the median relative error of mitigation over the sampled circuits \[\bar{\epsilon} = {\rm{med}}_{i=1}^{C}[ \epsilon_i(O, n,d) ].\]
\end{enumerate}
The choice of circuit classes and sampling methods are flexible and we discuss them in detail in the following section. 
In our benchmarks we will consider global observables, acting non-trivially on each qubit.
Measurements of global observables are typically affected by all errors occurring throughout a circuit whereas a local observable is affected by errors within its light-cone. Indeed, if the noisy operations outside the light-cone are described by completely positive trace-preserving maps, then their action on the identity observable cancels out. This assumption may fail if, for example, correlated errors across the cone partition occur. Therefore the choice of a global observable will place stronger constraints on depths/qubit number at which error mitigation gives improved estimators.  One might also consider observables with a fixed locality constraint determined by applications of interest.

There are two motivations behind the use of medians as a measure of overall success. First, we discussed how finite sampling statistics can give a long-tailed distribution of the relative error of mitigation. Secondly we observe for multiple experiments a similar long-tailed behaviour for the distribution over different circuits in the same class $\mathcal{C}(n,d)$.

 
\subsection{Circuits}
\label{sec:benchmark circuits}

\begin{figure*}
    \includegraphics[width=\textwidth]{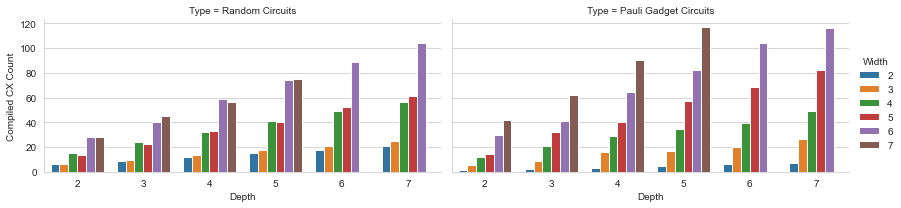}
    \caption{\textbf{\CX{} gate count of circuits used in the experiments of
    \cref{sec:results}.} These circuits are not mirrored. Compilation is onto
    the \texttt{ibm\_lagos} device, as discussed in \cref{sec:emulator}. Bars
    give the mean CX gate count over the circuits.}
    \label{fig:benchmark circuits count}
\end{figure*}

In this section, we formally define and motivate the circuits used in
benchmarking error mitigation methods. In particular, for each circuit class we
define a primitive \emph{layer} type, providing pseudocode generating it. A
layer acts on $n$ qubits, where $n$ is called the \emph{qubit number} of the circuit.
We specify the depth of a layer, and discuss how layers are composed to create
a \emph{depth} $d$ circuit from the given class. This terminology corresponds
to that used in \cref{sec:benchmark volume}. While the qubit number and depth are
indicative of the circuit sizes, the exact number of gates will vary between
circuit classes. The compiled circuits will additionally depend on the
architecture of the backend. This is discussed and displayed in
\cref{fig:benchmark circuits count}. The particular implementation generating
the circuits used in our experiments is available as specified in the
\hyperlink{code}{Code Availability}.

The circuit classes that we make use of are \emph{\random{}}, introduced in
\cref{sec:benchmark circuits random} and inspired by those used for quantum
volume experiments \cite{cross2018validating, mills2020applicationmotivated},
and \pauli{}, introduced in \cref{sec:benchmark circuits pauli} and inspired by
ansatz circuits for NISQ algorithms \cite{peruzzo2014variational,
barkoutsos2018quantum}.  Collectively, this selection of circuit classes
encompass several important potential applications of quantum computing,
covering circuits of varied depth, connectivity, and gate types, strengthening
the predictive power of our benchmarks. 

While these circuit classes allow us to draw conclusions about the performance
of error mitigation on specific applications, there are many circuits that
would not be covered by such classes. However, the mirroring method discussed
in \cref{sec:benchmark circuits mirror}, and the experiment design in
\cref{sec:experiments} are sufficiently general to allow for the use of
different classes of circuits, as may be desirable to others. General purpose
benchmark suites may provide inspiration for such choices
\cite{lubinski2023applicationoriented}.

We avoid favouring any device in particular by the design of the benchmark
circuits used here. The connectivity and the gate-set assumed by
the circuits is very general to ensure we benchmark the error mitigation
schemes as general purpose schemes. A compilation step will be required to
execute these circuits on a real device. The compilation strategy used may also
influence the performance of the error mitigation scheme. While the same
compilation strategy is used during error mitigated runs as with noisy runs,
compilation passes may have different effects on error mitigated and unaltered
circuits. As such we fix the compilation scheme used during our benchmarks to
avoid this dependency.
Note in particular that caution should be taken in the
case of mirrored circuits (see \cref{sec:benchmark circuits mirror}) to ensure that the circuits are not compiled to the identity.
 
 
\subsubsection{\titlecap{\random{}}}
\label{sec:benchmark circuits random}

While circuits required for applications are typically not random, sampling
from the output distributions of random circuits built from two-qubit gates has
been suggested as a means to demonstrate \supremacy{} \cite{bouland2018quantum,
movassagh2018efficient, boixo2018characterizing, aaronson2016complexity}.  By
utilising uniformly random two-qubit unitaries and all-to-all connectivity,
\emph{\random{}}, introduced now, provide a comprehensive benchmark.
 
The circuits used here -- taken from Ref.~\cite{mills2020applicationmotivated} and
similar to those used for the quantum volume benchmark
\cite{cross2018validating} -- are generated according to \cref{alg:benchmark
circuits random}, and are illustrated in \cref{fig:benchmark circuits random}. This circuit class consist of $d$ \emph{layers} of two-qubit gates acting between a bipartition of the qubits. 

\begin{figure}
	\begin{algorithm}[H]
		\caption{The pattern for building \random{}.}
		\label{alg:benchmark circuits random}
		\inout{Width, $n \in \mathbb{Z}$, depth, $d \in
			\mathbb{Z}$ and $\text{mirrored} \in \curlbrac{\text{True}, \text{False}}$}{Circuit, $C_n$}
		\begin{algorithmic}[1]
			\State Initialise n qubits, labelled $q_1 , ... , q_n$, to $\ket{0}$
			\State
			\If{mirrored}
			\State $d = \floor{\frac{d}{2}}$
			\EndIf
			\For{each layer $t$ up to depth $d$}
			\State \Comment This loop's content
			constitutes a \emph{layer}. 
			\State
			\State Divide the qubits into $\floor{\frac{n}{2}}$
			pairs $\curlbrac{q_{i,1}, q_{i,2}}$ at random.
			\State
			\ForAll{$i \in \mathbb{Z}$, $0 \leq i \leq
				\floor{\frac{n}{2}}$}
			\State
			\State Generate $\U{}_{i,t} \in \mathrm{SU} \brac{4}$
			uniformly at random according to the Haar measure.
			\State Enact the gate corresponding to the unitary
			$\U{}_{i,t}$ on qubits $q_{i, 1}$ and $q_{i,2}$.
			\State \Comment Decompositions of this gate can be
			found in \cite{tucci2005introduction}
			\If{mirrored}
			\State Enact the gate corresponding to the unitary
			$\U{}_{i,t}^{\dagger}$ on qubits $q_{i, 1}$ and $q_{i,2}$.
			\EndIf
			\State
			\EndFor
			\State
			\EndFor
			\State
			\State Measure all qubits in the computational basis.
		\end{algorithmic}
	\end{algorithm}
\end{figure}

\begin{figure*}
    \begin{subfigure}[t]{0.55\textwidth}
    \centering
	\begin{tikzpicture}[scale=0.8]
        \draw[very thick] (-0.5,1) node[anchor=east] {$q_n = \ket{0}$} -- (0,1);
        \node at (-0.5, 2) {\vdots};
		\draw[very thick] (-0.5,3) node[anchor=east] {$q_3 = \ket{0}$} -- (0,3);
		\draw[very thick] (-0.5,4) node[anchor=east] {$q_2 = \ket{0}$} -- (0,4);
		\draw[very thick] (-0.5,5) node[anchor=east] {$q_1 = \ket{0}$} -- (1.5,5);

		\draw[very thick] (0,0.6) rectangle (1,4.4) node[pos=.5, rotate=90] {$\U{}_{0,0}$};
		\draw[very thick] (1.5,2.6) rectangle (2.5,5.4) node[pos=.5, rotate=90]
        {$\U{}_{\floor{\frac{n}{2}},0}$};
		
		\draw[very thick, dotted] (1,3) -- (1.5,3);
		\draw[very thick, dotted] (1,4) -- (1.5,4);

		\draw[very thick] (2.5,5) -- (3,5);
		\draw[very thick] (2.5,4) -- (3,4);
		\draw[very thick] (2.5,3) -- (3,3);
		\draw[very thick] (1,1) -- (3,1);

        \draw[very thick, dotted] (3,5) -- (3.5,5);
		\draw[very thick, dotted] (3,4) -- (3.5,4);
		\draw[very thick, dotted] (3,3) -- (3.5,3);
		\draw[very thick, dotted] (3,1) -- (3.5,1);

        \draw[very thick] (3.5,5) -- (5.5, 5);
		\draw[very thick] (3.5,4) -- (5.5, 4);
		\draw[very thick] (3.5,3) -- (4, 3);
		\draw[very thick] (3.5,1) -- (4, 1);
		
		\draw[very thick] (4,0.6) rectangle (5,3.4) node[pos=.5, rotate=90]
        {$\U{}_{0,n}$};
		\draw[very thick] (5.5,3.6) rectangle (6.5,5.4) node[pos=.5, rotate=90]
        {$\U{}_{\floor{\frac{n}{2}},n}$};

        \draw[very thick] (6.5,5) -- (7, 5);
		\draw[very thick] (6.5,4) -- (7, 4);
		\draw[very thick] (5,3) -- (7, 3);
		\draw[very thick] (5,1) -- (7, 1);

		\draw[thick] (7,0.6) rectangle (8,1.4);
		\draw (7.1,0.9) .. controls (7.3,1.2) and (7.7,1.2) ..
		(7.9,0.9);
		\draw[thick, ->] (7.5, 0.8) -- (7.8, 1.3);

        \node at (7.5, 2) {\vdots};

		\draw[thick] (7,2.6) rectangle (8,3.4);
		\draw (7.1,2.9) .. controls (7.3,3.2) and (7.7,3.2) ..
		(7.9,2.9);
		\draw[thick, ->] (7.5, 2.8) -- (7.8, 3.3);

		\draw[thick] (7,3.6) rectangle (8,4.4);
		\draw (7.1,3.9) .. controls (7.3,4.2) and (7.7,4.2) ..
		(7.9,3.9);
		\draw[thick, ->] (7.5, 3.8) -- (7.8, 4.3);

		\draw[thick] (7,4.6) rectangle (8,5.4);
		\draw (7.1,4.9) .. controls (7.3,5.2) and (7.7,5.2) ..
		(7.9,4.9);
		\draw[thick, ->] (7.5, 4.8) -- (7.8, 5.3);
		
	\end{tikzpicture}
    \caption{\textbf{ An example of \random{}, as generated by
    \cref{alg:benchmark circuits random}.} }
    \end{subfigure}
    \hfill
    \begin{subfigure}[t]{0.4\textwidth}
    \centering
		\begin{tikzpicture}[scale =0.8]
		\draw[very thick] (-0.5,1) node[anchor=east] {$q_n = \ket{0}$} -- (0,1);
		\node at (-0.5, 2) {\vdots};
		\draw[very thick] (-0.5,3) node[anchor=east] {$q_3 = \ket{0}$} -- (0,3);
		\draw[very thick] (-0.5,4) node[anchor=east] {$q_2 = \ket{0}$} -- (0,4);
		\draw[very thick] (-0.5,5) node[anchor=east] {$q_1 = \ket{0}$} -- (0,5);
		
		\draw[very thick] (0,0.6) rectangle (1,5.4) node[pos=.5, rotate=90] {$\exp \brac{i\bigotimes_j s^0_j \alpha^0}$};
		
		\draw[very thick] (1,1) -- (1.5,1);
		\draw[very thick] (1,3) -- (1.5,3);
		\draw[very thick] (1,4) -- (1.5,4);
		\draw[very thick] (1,5) -- (1.5,5);
		
		\draw[very thick, dotted] (1.5,1) -- (2,1);
		\draw[very thick, dotted] (1.5,3) -- (2,3);
		\draw[very thick, dotted] (1.5,4) -- (2,4);
		\draw[very thick, dotted] (1.5,5) -- (2,5);
		
		\draw[very thick] (2,1) -- (2.5,1);
		\draw[very thick] (2,3) -- (2.5,3);
		\draw[very thick] (2,4) -- (2.5,4);
		\draw[very thick] (2,5) -- (2.5,5);
		
		\draw[very thick] (2.5,0.6) rectangle (3.5,5.4) node[pos=.5, rotate=90] {$\exp \brac{i\bigotimes_j s^d_j \alpha^d}$};
		
		\draw[very thick] (3.5,1) -- (4,1);
		\draw[very thick] (3.5,3) -- (4,3);
		\draw[very thick] (3.5,4) -- (4,4);
		\draw[very thick] (3.5,5) -- (4,5);
		
		\draw[thick] (4,0.6) rectangle (5,1.4);
		\draw (4.1,0.9) .. controls (4.3,1.2) and (4.7,1.2) ..
		(4.9,0.9);
		\draw[thick, ->] (4.5, 0.8) -- (4.8, 1.3);
		
		\node at (4.5, 2) {\vdots};
		
		\draw[thick] (4,2.6) rectangle (5,3.4);
		\draw (4.1,2.9) .. controls (4.3,3.2) and (4.7,3.2) ..
		(4.9,2.9);
		\draw[thick, ->] (4.5, 2.8) -- (4.8, 3.3);
		
		\draw[thick] (4,3.6) rectangle (5,4.4);
		\draw (4.1,3.9) .. controls (4.3,4.2) and (4.7,4.2) ..
		(4.9,3.9);
		\draw[thick, ->] (4.5, 3.8) -- (4.8, 4.3);
		
		\draw[thick] (4,4.6) rectangle (5,5.4);
		\draw (4.1,4.9) .. controls (4.3,5.2) and (4.7,5.2) ..
		(4.9,4.9);
		\draw[thick, ->] (4.5, 4.8) -- (4.8, 5.3);
		
	\end{tikzpicture}
    \caption{\textbf{ An example of \pauli{}, as generated by
    \cref{alg:benchmark circuits pauli}.} }
    \end{subfigure}
    \caption{\textbf{Example circuits.}
    These particular examples have mirrored set to false. In the case of mirrored \pauli{}, one can imagine appending the gate $\exp \brac{-i\bigotimes_j
s^t_j \alpha^t}$ after the gate $\exp \brac{i\bigotimes_j s^t_j
\alpha^t}$. Respectively in the case of mirrored \random{}, one can imagine
appending the gate $\U{}_{i,t}^{\dagger}$ after the gate $\U{}_{i,t}$. Note that the gates $\U{}_{i,t}$ in the random circuit class act between 2 randomly selected qubits.}
    \label{fig:benchmark circuits random}
\end{figure*}

By utilising uniformly random two-qubit unitaries, \random{} test the ability
of the error mitigation scheme to mitigate errors acting on a universal gate set.
Allowing two-qubit gates to act between any pair of qubits in the uncompiled
circuit, means \random{} avoid favouring any device in particular. This choice
adheres closely to our motivations of being hardware-agnostic. Because of this generality \random{} are complementary to the other classes introduced in this work.
 

\subsubsection{\titlecap{\pauli{}}}
\label{sec:benchmark circuits pauli}

Pauli gadgets \cite{cowtan2019phase} are quantum circuits implementing an
operation corresponding to exponentiating a Pauli tensor. Sequences of Pauli
gadgets acting on qubits form \emph{product formula} circuits, most commonly
used in Hamiltonian simulation \cite{Berry2007}. Many algorithms employing
these circuits require fault-tolerant devices, but they are also the basis of
trial state preparation circuits in many variational algorithms, which are the
most promising applications of noisy quantum computers. 

A notable example of this in quantum chemistry is the physically-motivated
\emph{Unitary Coupled Cluster} family of trial states used in the variational quantum
eigensolver (VQE) \cite{peruzzo2014variational, barkoutsos2018quantum}.  
Motivated by these practical applications, we include in our benchmarks circuits with a similar structure.
The \pauli{} are built as in \cref{alg:benchmark circuits pauli}, and may be visualised as in
\cref{fig:benchmark circuits random}. They are constructed from several layers
of Pauli Gadgets, each acting on a random subset of $n$ qubits. Note that the
circuits in this class differ from running end-to-end VQE. Focusing on the
state preparation portion of a VQE circuit, we might deduce performance of
error mitigation when running the VQE on ansatze of similar type.

\begin{figure}
	\begin{algorithm}[H]
		\caption{The pattern for building \pauli{}.}
		\label{alg:benchmark circuits pauli}
		\inout{Width, $n \in \mathbb{Z}$, depth, $d \in \mathbb{Z}$ and
			$\text{mirrored} \in \curlbrac{\text{True}, \text{False}}$}
		{Circuit, $C_n$}
		\begin{algorithmic}[1]
			\State Initialise $n$ qubits, labelled $q_1 , ... , q_n$, to $\ket{0}$.
			\State
			\If{mirrored}
			\State $d = \floor{\frac{d}{2}}$
			\EndIf
			\For{each layer $t$ up to depth $d$}
			\State \Comment This loop's content
			constitutes a \emph{layer}. 
			\State
			\State Select a random string $s^t \in
			\curlbrac{\I{},\X{},\Y{},\Z{}}^n$
			\State Generate random angle $\alpha^t \in \sqrbrac{0 , 2
				\pi}$
			\State Enact the Pauli gadget corresponding to the unitary
			$\exp \brac{i\bigotimes_j s^t_j \alpha^t}$ on qubits $q_1 , ...
			, q_n$.
			\State \Comment Decompositions of this gate can be
			found in \cite{cowtan2019phase}
			\If{mirrored}
			\State Enact the Pauli gadget corresponding to the unitary
			$\exp \brac{-i\bigotimes_j s^t_j \alpha^t}$ on qubits $q_1 ,
			...  , q_n$.
			\EndIf
			\State
			\EndFor
			\State
			\State Measure all qubits in the computational basis
		\end{algorithmic}
	\end{algorithm}
\end{figure}

 		
\subsubsection{Mirrored Circuits} 
\label{sec:benchmark circuits mirror}

Besides the division by circuit class, given by the specific layer type,
circuits are also defined as being mirrored or un-mirrored. While un-mirrored
circuits consist of independently generated random layers of a fixed type,
mirrored circuits consist of randomly generated layers, each followed by their
inverse. Mirrored circuits correspond to an identity operation and thus have
the property that their expectation values can be efficiently computed with
increasing qubit number and depth.

Note that Mirrored circuits in this work are inspired by, and are a special
case of, those introduced in \cite{measuring2021proctor}. ``Mirroring'' in
\cite{measuring2021proctor} consists of concatenating a circuit with its
quasi-inverse (its inverse up to Pauli operations) and inserting Pauli
operations before, after and between the circuit and its quasi-inverse. The
ideal behaviour of such a circuit would be to produce the input state, inspired
by the Loschmidt echo. In our case we take the quasi inverse to be the inverse,
and do not add layers of Pauli operations.
 	 
In both \cref{alg:benchmark circuits random} and \cref{alg:benchmark circuits
pauli} the option to generate mirrored circuits is provided as an input
parameter. This has the effect of inverting each layer of the circuit once the
layer has been applied. Random layers are generated and inverted a total of
$\floor{\frac{d}{2}}$ times to ensure that the final depth is $d$, assuming
that $d \in 2\mathbb{Z}$. This results in a circuit implementing the identity, with ideal expectation value that depends only on the observable and input state. Therefore, mirrored circuits can provide scalable benchmarks to assess the performance of error mitigation strategies. We discuss this further in \cref{app:certify}.

 
\subsubsection{Experiment Design}
\label{sec:experiments}
In our experiments, for a given qubit number and depth we choose a fixed number of random instances from each class of Random  and Pauli-Gadget circuits and their mirrored versions. The measurements are in the computational basis with the target observables set to $O = \bigotimes_{i=0}^{n} \Z_i$ and input state $\rho_{0} = |0\>^{\otimes n}$. The choice of a global observable was motivated in \cref{sec:benchmark volume}.


From the randomly generated circuit instances we select those with ideal expectation values within a fixed range (specifically we select values from $0.4$ to $0.6$).  As the circuit size increases the probability of generating circuits with expectation values in this range falls exponentially. This limits the size of such circuits that we can generate, demonstrating an additional advantage of the mirrored circuits which in this case have a fixed ideal expectation value 1.

We require this control of the ideal expectation value for the following
reason:  Random circuits show anti-concentration properties with increasing
depth, meaning that for Pauli observables the expectation values will be close
to zero with high probability\cite{dalzell2022random}. However, our
experiments, particularly the hardware runs, are limited by finite size
sampling so that the obtained data ${\bf{D}}= \{D^{q}_{i,m}\}_{i,m}$ incurs a
variance depending on the fixed number of shots.  Fitting this data to the
surface $\mathfrak{F}({\bf{D}}, \<\hat{O}\>_{EM})= 0$ to determine an error
mitigated estimator can be numerically unstable if the variance due to finite
sampling is comparable to the value of the noisy data $\bf{D}$ itself. This is
particularly relevant for very low ideal expected values since accumulation of
noise within the circuit results in a further decreased value (typically
dropping exponentially with number of noisy gates). To avoid this situation,
where the error mitigation performance decrease with increasing depth
can be (partly) attributed to the random generation of benchmark circuits, we
restrict to high/fixed range expected values.  We refer to
\cref{ap:finite-sampling-fit} for further details.


 
\section{Results}
\label{sec:results}

\newcommand{\includeplot}[1]{\includegraphics[height=4.5cm]{imag/#1.png}}

\begin{figure*}[t!]
	\centering

    \includegraphics[width=5cm]{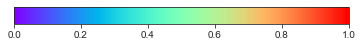}

	\hfill
	\begin{subfigure}[t]{0.25\textwidth}
		\centering
		\includeplot{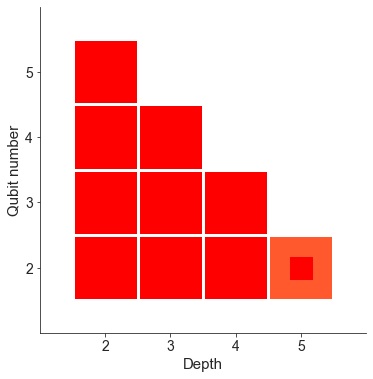}
		\caption{\texttt{ibm\_lagos}, ZNE, \pauli{}.}
        \label{fig:results pauli lagos false zne}
	\end{subfigure}
	\hfill
	\begin{subfigure}[t]{0.25\textwidth}
		\centering
		\includeplot{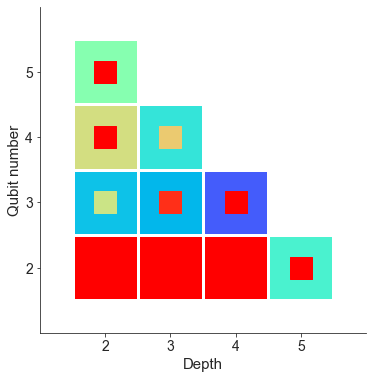}
		\caption{\texttt{ibm\_lagos}, CDR, \pauli{}.}
        \label{fig:results pauli lagos false cdr}
	\end{subfigure}
	\hfill
	\begin{subfigure}[t]{0.2\textwidth}
		\centering
		\includeplot{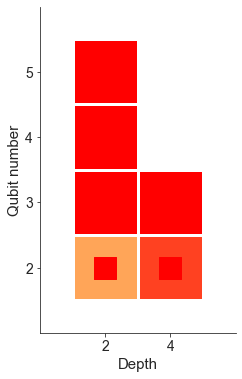}
		\caption{\texttt{ibm\_lagos}, ZNE, Mirrored \pauli{}.}
        \label{fig:results pauli lagos true zne}
	\end{subfigure}
	\hfill
	\begin{subfigure}[t]{0.2\textwidth}
		\centering
		\includeplot{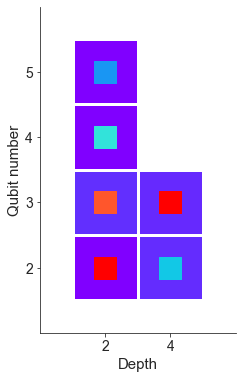}
		\caption{\texttt{ibm\_lagos}, CDR, Mirrored \pauli{}.}
        \label{fig:results pauli lagos true cdr}
	\end{subfigure}
	\hfill

	\hfill
	\begin{subfigure}[t]{0.25\textwidth}
		\centering
		\includeplot{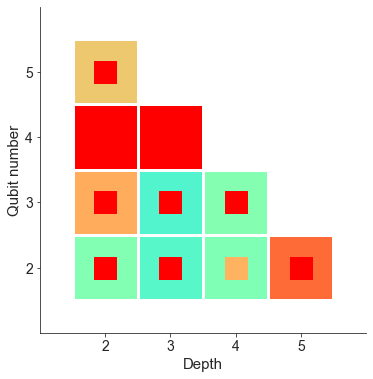}
		\caption{\texttt{ibmq\_casablanca}, ZNE, \pauli{}.}
        \label{fig:results pauli casablanca false zne}
	\end{subfigure}
	\hfill
	\begin{subfigure}[t]{0.25\textwidth}
		\centering
		\includeplot{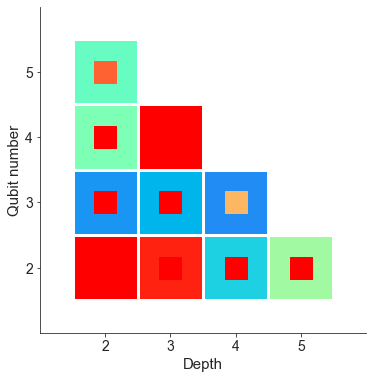}
		\caption{\texttt{ibmq\_casablanca}, CDR, \pauli{}.}
        \label{fig:results pauli casablanca false cdr}
	\end{subfigure}
	\hfill
	\begin{subfigure}[t]{0.2\textwidth}
		\centering
		\includeplot{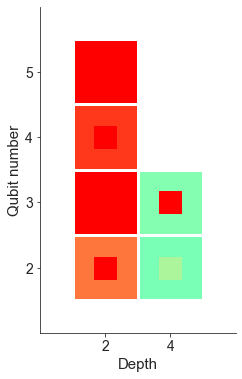}
		\caption{\texttt{ibmq\_casablanca}, ZNE, Mirrored \pauli{}.}
        \label{fig:results pauli casablanca true zne}
	\end{subfigure}
	\hfill
	\begin{subfigure}[t]{0.2\textwidth}
		\centering
		\includeplot{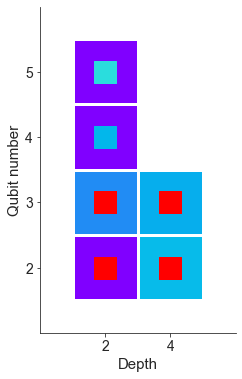}
		\caption{\texttt{ibmq\_casablanca}, CDR, Mirrored \pauli{}.}
        \label{fig:results pauli casablanca true cdr}
	\end{subfigure}
	\hfill
    \caption{\textbf{Volumetric plots for performance of error mitigation on
    \pauli{}.} \texttt{ibm\_lagos} and
    \texttt{ibmq\_casablanca} are used with ZNE or CDR on \pauli{}
    and mirrored \pauli{}.  Median $\bar{\epsilon}$ (outer square) and worst-case (inner
    square) relative error of mitigation over the sampled circuits are shown,
    metric that ranges between $(0,1)$ for successful mitigation with lower
    values corresponding to better performance. Each square corresponds to 5
    sampled circuits of the specified type with a total $1.6\times10^{5}$
    total shot budget per each mitigated expectation value. }
	\label{fig:pauli-hardware}
\end{figure*}

\begin{figure*}[t!]
	\centering

    \includegraphics[width=5cm]{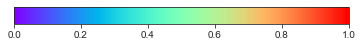}

	\hfill
	\begin{subfigure}[t]{0.25\textwidth}
		\centering
		\includeplot{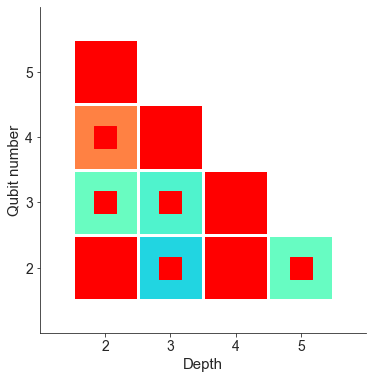}
		\caption{\texttt{ibm\_lagos}, ZNE, random{}.}
        \label{fig:results random lagos false zne}
	\end{subfigure}
	\hfill
	\begin{subfigure}[t]{0.25\textwidth}
		\centering
		\includeplot{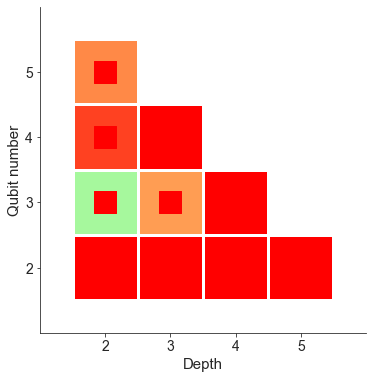}
		\caption{\texttt{ibm\_lagos}, CDR, \random{}.}
        \label{fig:results random lagos false cdr}
	\end{subfigure}
	\hfill
	\begin{subfigure}[t]{0.2\textwidth}
		\centering
		\includeplot{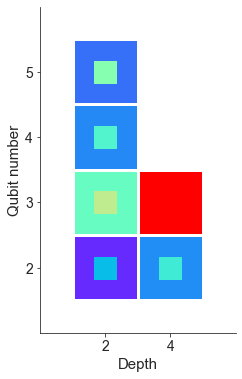}
		\caption{\texttt{ibm\_lagos}, ZNE, Mirrored \random{}.}
        \label{fig:results random lagos true zne}
	\end{subfigure}
	\hfill
	\begin{subfigure}[t]{0.2\textwidth}
		\centering
		\includeplot{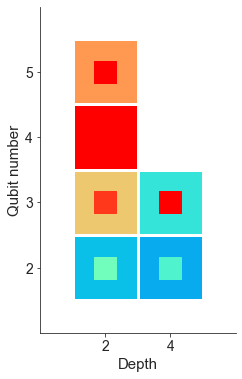}
		\caption{\texttt{ibm\_lagos}, CDR, Mirrored \random{}.}
        \label{fig:results random lagos true cdr}
	\end{subfigure}
	\hfill

	\hfill
	\begin{subfigure}[t]{0.25\textwidth}
		\centering
		\includeplot{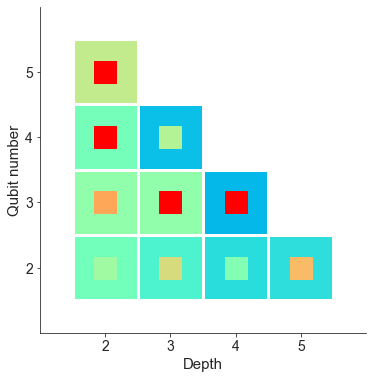}
		\caption{\texttt{ibmq\_casablanca}, ZNE, \random{}.}
        \label{fig:results random casablanca false zne}
	\end{subfigure}
	\hfill
	\begin{subfigure}[t]{0.25\textwidth}
		\centering
		\includeplot{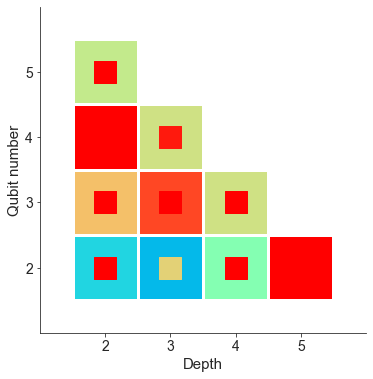}
		\caption{\texttt{ibmq\_casablanca}, CDR, \random{}.}
        \label{fig:results random casablanca false cdr}
	\end{subfigure}
	\hfill
	\begin{subfigure}[t]{0.2\textwidth}
		\centering
		\includeplot{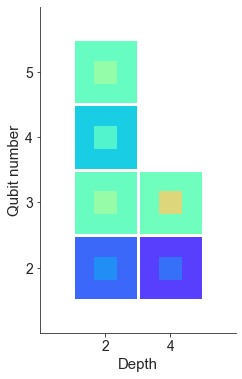}
		\caption{\texttt{ibmq\_casablanca}, ZNE, Mirrored \random{}.}
        \label{fig:results random casablanca true zne}
	\end{subfigure}
	\hfill
	\begin{subfigure}[t]{0.2\textwidth}
		\centering
		\includeplot{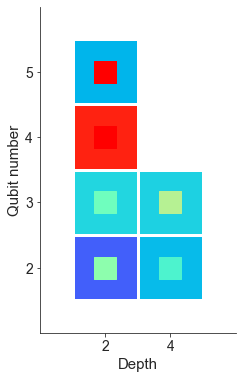}
		\caption{\texttt{ibmq\_casablanca}, CDR, Mirrored \random{}.}
        \label{fig:results random casablanca true cdr}
	\end{subfigure}
	\hfill
    \caption{\textbf{Volumetric plots for performance of error mitigation on
    \random{}.} \texttt{ibm\_lagos} and
    \texttt{ibmq\_casablanca} are used with ZNE or CDR on \random{}
    and mirrored \random{}.  Median $\bar{\epsilon}$ (outer square) and worst-case (inner
    square) relative error of mitigation over the sampled circuits are shown,
    metric that ranges between $(0,1)$ for successful mitigation with lower
    values corresponding to better performance. Each square corresponds to 5
    sampled circuits of the specified type with a total $1.6\times10^{5}$
    total shot budget per each mitigated expectation value. }
	\label{fig:random-hardware}
\end{figure*}

We use the benchmarks introduced in \cref{sec:benchmark} to assess the
performance of ZNE and CDR on both superconducting hardware and noisy simulator
backends. In \cref{fig:pauli-hardware} and \cref{fig:random-hardware} we
present volumetric plots displaying the relative error of mitigation for
\pauli{} and \random{} respectively, when run on both \texttt{ibm\_lagos} and
\texttt{ibmq\_casablanca}. Extensive noisy simulations are provided in the
\cref{ap:noise-model}.

\subsection{Noisy Classical Simulations}
\label{sec:results classical}

The first set of benchmarks employ a local depolarising noise model with error rate of $e_1 =10^{-3}$ and $e_2 = 10^{-2}$ for single and two qubit gates respectively.\footnote{To create this model we make use of the qiskit noise
	model module, as described at
	\url{https://qiskit.org/documentation/apidoc/aer_noise.html}. For more details
	see \hyperlink{code}{Code Availability}.} 
These values are chosen to be broadly inline with those reported by several hardware providers, but do not correspond to any device in particular. In this case there is no restriction on the pairs of qubits between which two qubit gates can act. As such only minimal compilation to the simulator's gate set is required.

The results of these experiments are presented in the volumetric plots of \cref{fig:results
depolarising square,fig:results depolarising deep}. For each fixed width/depth (i.e square) 10 circuits are sampled from the specified class, with $5\times 10^{5}$ shots per circuit for each error mitigation protocol, and $10^5$ in the case of the noisy expected value. This shot count is higher than could be taken from real backends at our disposal in a reasonable runtime. This high shot count is used here to present an idealised scenario and illustrate boundaries of qubit count and circuit depth beyond which error mitigation fails to improve the noisy estimator of the expected value. In particular, such boundaries align with the theoretical analysis limiting the depth to scale with a constant inverse of the error rate $ d = O(1/e_2)$ \cite{cai2021practical,cai2021multiexponential}.

In the case of ZNE, we use an exponential fit and the target circuit is folded repeatedly for a number of $\lambda= 1, 3, 5, 7, 9$ times. The shot budget is distributed equally between these noise scaling values. Curve fitting is achieved via a least-square optimisation. For CDR, we distribute the total shot budget equally to each of the $21$ required circuits; namely $20$ near-Clifford circuits, and $1$ unaltered circuit. Generation of the training set fixes the number of non-Clifford operations to $10$ and uniformly at random generates the new near-Clifford circuits by pair replacements from an initial seed circuit. In the case where the total number of non-Clifford gates is less than $10$, at least one non-Clifford gate is replaced at random. Additionally, using the classical simulation of these circuits we test that the training data gives a well-conditioned matrix for the linear regression to be performed and generate new training circuits if that is not the case.

\paragraph{There are circuit sizes where ZNE and CDR are consistently
beneficial} Comparing first the performance of CDR and ZNE, we note from
\cref{fig:results depolarising square false zne,fig:results depolarising square
false cdr} that for \random{} in the size ranges explored, both show an
improvement over the case where error mitigation is not used. ZNE appears to
perform particularly well in this domain, outperforming CDR for each circuit
size explored. To identify the limits of the utility of CDR and ZNE we extend
the depth of the circuit explored using mirrored circuits, as discussed in
\cref{sec:benchmark circuits}. We see from \cref{fig:results depolarising
square true zne,fig:results depolarising square true cdr} that while ZNE
outperforms CDR for smaller circuit sizes, the range of circuit sizes where CDR
outperforms the use of no error mitigation appears to be larger than for ZNE.

\paragraph{CDR outperforms ZNE with \pauli{}} Comparing \cref{fig:results
depolarising deep} to \cref{fig:results depolarising square} we note that while
ZNE outperform CDR in case of \random{}, this appears not to be the case for
\pauli{}. As shown in \cref{fig:benchmark circuits count}, the \CX{} gate count
of circuits with a fixed qubit number and depth are similar between the
\random{} and \pauli{} classes. As such this improved relative performance of
CDR is likely because there are a smaller proportion of non-Clifford gates in
\pauli{} than \random{}. In the case where there are few non-Clifford circuits,
the training circuits are little different from the original circuit, which
improved the accuracy of the CDR training procedure.

\paragraph{More refined device modelling reduces performance} The second set of
benchmarks employ device emulation that includes many properties of the
corresponding real backend used in \cref{sec:results real}.\footnote{To achieve
device emulation use the pytket \textsf{IBMQEmulatorBackend} as described at
\url{https://cqcl.github.io/pytket-extensions/api/qiskit/api.html}.  For more
details see \hyperlink{code}{Code Availability}.} These properties include
connectivity of the architecture, available gate set, and an expansion of the
noise model to include thermal relaxation and readout errors, in addition to a
local depolarising model with parameters corresponding to calibration results.
Exact values of these properties can be found in \cref{sec:emulator}. The
parameters used for ZNE and CDR are identical in this case to those used when
performing depolarising noise simulation, with the exception that the total
shot budget is reduced to $1.6\times 10^{5}$ for the mitigated experiment, and
$0.32 \times 10^5$ for the unmitigated experiments. The shot budget is reduced
to match that of the real device experiments, where computing resources are
more scarce. 
Comparing \cref{fig:results depolarising square false zne,fig:results
depolarising square true zne,fig:results depolarising deep false
zne,fig:results depolarising deep true zne} to \cref{fig:results emulator
square false zne,fig:results emulator square true zne,fig:results emulator deep
false zne,fig:results emulator deep true zne} we see a marked fall in the
performance of ZNE when additional device constraints are impose by emulation.

\subsection{Quantum Hardware Backend}
\label{sec:results real}

Finally, we perform volumetric benchmarks to assess and compare the performance of ZNE and CDR error mitigation on \texttt{ibmq\_lagos} and \texttt{ibmq\_casablanca} as measured by the relative error of mitigation.  

The experiments used in the volumetric benchmarks are produced according to \cref{sec:benchmark circuits} and employ the two types of circuit classes as described in \cref{sec:experiments}. For each fixed qubit number and depth we use $5$ circuits on which we sequentially perform CDR, ZNE and no error mitigation. The same sets of circuits are used on both \texttt{ibmq\_casablanca} and \texttt{ibm\_lagos}. As discussed, the parameters used for ZNE and CDR are the same as in the case of the depolarising noise simulations, and the device emulation experiments. As in the case of the emulation experiments, the shot budget is set to $1.6 \times 10^5$ for the error mitigated experiments, and $0.32 \times 10^{5}$ for the unmitigated experiments.

It is important to emphasize that for any given circuit we run the two mitigation schemes, and the unmitigated experiments, back-to-back so that any time drift in the device's noise parameters do not skew the performance comparison. For both devices the calibration data during these experiments can be found in \cref{tab:backend properties casablanca} and \cref{tab:backend properties lagos}. Furthermore, they share the same architecture connectivity. 

\paragraph{Improvements are inconsistent} As seen
in \cref{fig:random-hardware,fig:pauli-hardware}, for a fixed qubit number and
depth (as measured by the number of layers from a specific circuit class), we
typically found at least one sampled circuit for which error mitigation does
not improve upon the noisy expectation value. This occurred for both error
mitigation methods even when their median performance gave a 2-fold or higher
error reduction (i.e $\bar{\epsilon} \leq 0.5$). This stands in contrast with
the emulation results, where there was a significant size region with different
$(n,d)$ for which error mitigation was successful even in the worst case.
Method parameters and shot budget were the same for emulation and device
experiments.

\paragraph{Improvements are less on \texttt{ibm\_lagos}} Strikingly, whilst the
dominating two qubit error rates are within 10\% for the two devices,
performance of error mitigation drastically decreased on \texttt{ibm\_lagos}.
In particular, as seen in \cref{fig:results pauli lagos false zne,fig:results
pauli lagos true zne}, the structured \pauli{} showed little to no improvement
over the noisy expectation value when exponential ZNE was performed.  This is
not due to \texttt{ibm\_lagos} producing results with low absolute error. As
indicated in \cref{tab:additional results pauli,tab:additional results random}
the absolute error of the results from \texttt{ibm\_lagos} are in fact
typically higher than that of \texttt{ibmq\_casablanca}.  Comparatively, as
seen in \cref{fig:results pauli casablanca false zne}, on
\texttt{ibmq\_casablanca} for small qubit sizes $n=3$ and $2$, ZNE mitigation
was successful with a median relative error of $0.51$ respectively $0.50$ for
$4$ Pauli circuit layers.  However, comparing \cref{fig:results pauli
casablanca false cdr,fig:results pauli casablanca true cdr} against
\cref{fig:results pauli lagos false cdr,fig:results pauli lagos true cdr}, we
see that the performance of CDR was consistent between the two devices for this
circuit class, and on average scaled well to the largest circuit sizes
considered. 

On the other hand, we see from \cref{fig:random-hardware} that for Random
circuits both ZNE and CDR had worse performance on \texttt{ibm\_lagos} compared
to \texttt{ibmq\_casablanca}, with the limits $\bar{\epsilon} >1$ reached
within the sizes considered in both depth and qubit number. For this circuit
class, as seen in \cref{fig:results random casablanca false zne,fig:results
random casablanca true zne}, 
the performance of ZNE on \texttt{ibm\_casablanca} consistently reached on
average a reduction in the noisy estimator's absolute error by 2 or more for
all sizes considered. This was generally slightly better than CDR, although
both methods still improved the noisy estimators even for larger qubit sizes
and depths. 

As a result of these features, each appearing unique to the particular device,
we do not expect the results presented here to carry over generically to other
devices. This is particularly the case for architectures other than
superconducting.

\paragraph{Mirrored circuits are optimistic} In the case of mirrored circuits,
the performance was more consistent between the two devices. Notably, as seen
in \cref{fig:results pauli casablanca true cdr,fig:results pauli lagos true
cdr}, on mirrored Pauli gadget circuits CDR decreased, on average, the absolute
error in the noisy estimator by up to an order of magnitude for most circuit
sizes. This corresponds in \cref{fig:pauli-hardware} to a median relative error
of mitigation of $~0.1$ or less, and the results were similar for both devices.
Comparing \cref{fig:results pauli casablanca false zne,fig:results pauli lagos
false zne} against \cref{fig:results pauli casablanca true zne,fig:results
pauli lagos true zne} shows ZNE mirrored Pauli circuits were a good predictor
of the poor performance of non-mirrored circuits with similar dimensions.

Surprisingly, the relative error of mitigation for mirrored random circuits
improved from $0.45$ on \texttt{ibmq\_casablanca} to $0.15$ on
\texttt{ibm\_lagos} for $n = 5$ qubits, as seen in \cref{fig:results random
casablanca true zne} and \cref{fig:results random lagos true zne},
respectively. This suggests that while mirrored circuits provide scalable
benchmarks to assess performance of error mitigation, they generally
overestimate the median relative error for circuits of similar sizes using both
of the methods considered. As such they may be taken to provide an prediction
of an upper bound on the performance of error mitigation schemes, and so
indicating when a scheme should be expected to bring no benefit.


 \section{Conclusions and Outlook}
 \label{sec:conclusion} 

In this work we introduce Qermit -- an open source package for error mitigation with a composable software architecture that allows for combining different error mitigation methods and facilitates development of new techniques.

A key question we address is \emph{given a particular device and class of
circuits, which error mitigation has, on average, a better performance?} To
that aim, we design volumetric benchmarks to assess the performance of error
mitigation methods for a given class of quantum circuits with varying qubit
number and depth. The methodology is based on frameworks for benchmarking
quantum devices \cite{BlumeKohout2020volumetricframework}. We use this
framework here to delineate the boundary beyond which a given error mitigation
technique fails on average to give an improved estimator of the target
expectation value. 

In this setup, we compare the performance of ZNE and CDR implementations on current superconducting hardware. We show how noise in real devices places even stronger constraints on the scalability of error mitigation than theoretical lower bounds as for example derived in \cite{takagi2021fundamental} or determined from classical noisy simulations \cite{bultrini2021unifying}. Qualitatively, we found that even for a low total shot budget ($\approx 10^5$ per mitigated expected value) CDR  generally outperforms ZNE on structured Pauli circuits and slightly underperforms on random SU(4) circuits. This generic behaviour is captured by the emulation which includes depolarisation, thermal relaxation and readout in the noise model as well as the device's architecture. However, emulation largely overestimates the performance with successful mitigation for significantly larger circuit sizes compared to those accessible with real hardware. In contrast, a simplified depolarising noise model without connectivity constraints fails to predict even qualitative features of error mitigation performance. 

On real hardware, we find that for circuits of the same type and depth the
relative error of mitigation can vary extensively. Often, it results in
unsuccessful mitigation in the worst case even when on average the error in the
expectation value is reduced by an order of magnitude. We run the same
benchmarks on two superconducting devices with the same connectivity and
comparable calibration data and find a high level of variability in the
performance of both error mitigation methods. This was particularly striking
for ZNE, where the use of digitised method to artificially increase noise level
    may be more susceptible to the different noise profiles. These results flag
    two issues; first, that fine grained noise characterisation is needed to
    predict behaviour of error mitigation on hardware and second, that without
    access to the ideal expectation value one needs to validate that error
    mitigation has succeeded. 

Our benchmarks include mirrored circuits, which can be efficiently classically
simulated as the number of qubits grows. As such, mirrored circuits provide a
useful benchmark of the performance of error mitigation, particularly when
classical simulation of their un-mirrored counterparts is inaccessible. We
observe that, given a fixed qubit number, depth, shot budget and device,
mirrored circuits typically produced a lower average relative error of
mitigation for both ZNE and CDR compared to un-mirrored circuits from the same
class. This illustrates that mirrored circuits give an upper bound on the
relative error of mitigation.  This feature remains constant across devices,
circuit classes and schemes, and so it gives a way to infer when EM will not
produce improved results.


Due to the emphasis on composability, Qermit is particularly well suited to
explore the application of multiple noise tailoring and mitigation techniques
for the same circuit and observable. In future work we will explore the
performance of combined methods, which is particularly well motivated by recent
results that show significant improvements when using Pauli frame
randomisation, ZNE and dynamical decoupling in combination, and with additional
bespoke device calibration \cite{Kim2023, kim2021scalable}. In particular, for
these combined methods the error mitigated estimates of magnetisation gave a
better approximation than the leading approximate classical tensor network
simulations for 2D Ising models on up to $N=127$ qubits. Note that these
results do not contradict those of this paper as we do not explore here
combinations of error mitigation schemes, or techniques which make such
extensive use of device characteristics.  However, this indicates that Qermit's
composable design makes it suitable for testing how useful error mitigation can
be to achieving practical quantum advantage.

It would also be pertinent to understand how the performance of error
mitigation schemes varies between different device architectures, such as ion
traps devices. As we have seen, for example when comparing the classical
simulations of \cref{ap:noisy simulation} to those performed on a real device
in \cref{sec:results real}, idealised noise models can be a poor predictor of
real performance. This discrepancy will likely be emphasised by using different
devices, between which noise models vary greatly.  Even apparently similar
devices were shown in \cref{sec:results real} to perform differently, and so we
do not expect our results to generically carry over to other superconducting
devices. Qermit facilitates such experiments as it is implemented with pytket
\cite{Sivarajah_2020}, which gives access to an assortment of devices.  While
in \cref{ap:noisy simulation} we note some observations regarding the
performance of error mitigation with changing noise levels, a more refined and
extensive study with more complex noise models would be of great interest.

Since all parameters including total shot budget per mitigated expected value
are fixed, the discrepancy in performance between the two devices can be
attributed to the noise bias and different noise profiles.  Therefore, it would
be of interest to determine how the results of the volumetric benchmarks change
for the \emph{same} device at different calibration times. This was not
possible in the case of \texttt{ibm\_casablanca} as the machine was taken out
of use during our experiments, so we leave this type of assessment for future
work.

It will also be particularly interesting to explore the effect of finite
sampling in the case of ion traps, where the rate at which shots are generated
is lower. Such restrictions will impact the accuracy with which a functional
model, as discussed in \cref{sec:em framework}, can be approximated. This could
manifest, for example, in the accuracy of the learnt extrapolation function in
the case of ZNE, and so the variance in the learnt zero noise limit. In
\cref{ap:finite-sampling-fit} we introduce theoretical tools to explore this
particular point, inspired by numerical results revealing strong performance
variation depending on shot budget availability \cite{bultrini2021unifying}. We
look to future work to establish strictly how the relative error of mitigation
varies not just with circuit size as explored here, but also with shot count.

Finally, would be insightful to compare our approach, targeting an
understanding of the practical utility of error mitigation, to those exploring
information theoretic bounds to performance \cite{takagi2021fundamental,
quek2023exponentially, takagi2022universal, tsubouchi2023universal}. In such
results, lower bounds on the resources required to ensure beneficial use of
error mitigation are derived, and are shown to grow exponentially with circuit
size. We anticipate that the approach of our work will be complementary to
those results, firstly as a means of verifying the predicted scaling. Secondly
we imagine that our techniques could be used to explore more complicated
unknown noise models.

\hypertarget{code}{\paragraph{Data Availability}}

A complete set of results can be found in \cite{experimentresults}. This
additionally includes the circuits used, and other data required to reproduce
the experiments.

\hypertarget{qermit}{\paragraph{Qermit Installation and Usage}}

Information on the installation and use of Qermit is available in
the documentation at:
\begin{center}
    \url{www.qerm.it}
\end{center}
and the user manual at:
\begin{center}
    \url{https://cqcl.github.io/Qermit/manual/}
\end{center}
Qermit is available for the Mac, Linux and Windows operating systems
via PyPI. Qermit may be installed using the command line operation:
\begin{center}
    \texttt{pip install qermit}
\end{center}
The source code for Qermit is hosted at:
\begin{center}
    \url{https://github.com/CQCL/qermit}
\end{center}

 \paragraph{Acknowledgements}

Thanks to Steven Herbert and Hussain Anwar for insightful feedback. This work
was supported by Innovate UK Project No: 10001712. ``Noise Analysis and
Mitigation for Scalable Quantum Computation''. We acknowledge the use of IBM
Quantum services for this work. The views expressed are those of the authors,
and do not reflect the official policy or position of IBM or the IBM Quantum
team.

\printbibliography

\appendix

\section{Code Snippets}
\label{sec:code}

\subsection{\texttt{MitRes}}
\label{sec:code mitres}

The \texttt{MitRes.run} method takes a list of \texttt{CircuitShots} objects as
an argument. Each \texttt{CircuitShots} objects contains the basic information
required to run a circuits, namely; a state preparation circuit, and the number
of shots to run for the circuit.

The example code snippet:
\begin{lstlisting}[language=Python]
from qermit import MitRes, CircuitShots
from pytket import Circuit
from pytket.extensions.qiskit import AerBackend

mitres = MitRes(backend=AerBackend())
c = Circuit(2, 2).H(0).Rz(0.25, 0).CX(1, 0).measure_all()
results = mitres.run([CircuitShots(Circuit=c, Shots=50)])
print(results[0].get_counts())

graph = mitres.get_task_graph()
\end{lstlisting}
produces the output:
\begin{lstlisting}
Counter({(0, 0): 30, (1, 0): 20})
\end{lstlisting}
with \texttt{graph} being the \texttt{TaskGraph} of \cref{fig:qermit mitres taskgraph}.

\subsection{\texttt{MitEx}}
\label{sec:code mitex}

The \texttt{MitEx.run} method takes a list of \texttt{ObservableExperiment}
objects as an argument. Each \texttt{ObservableExperiment} objects contains the
basic information required to estimate the expectation value of an observable:
a state preparation circuit, a dictionary between symbols and parameter values
(where appropriate), a pytket \texttt{QubitPauliOperator} detailing the
operator being measured and used for preparing measurement circuits, and the
number of shots to run for each measurement circuit. The following is an
example of an \texttt{ObservableExperiment} set up.

\begin{lstlisting}[language=Python]
from qermit import AnsatzCircuit, ObservableExperiment, ObservableTracker, SymbolsDict
from pytket.pauli import Pauli, QubitPauliString
from pytket.utils import QubitPauliOperator
from pytket import Circuit, Qubit

qubit_pauli_string = QubitPauliString([Qubit(1), Qubit(2)], [Pauli.Z, Pauli.Z])
qubit_pauli_operator = QubitPauliOperator({qubit_pauli_string: 1.0})
ansatz_circuit = AnsatzCircuit(
    Circuit=Circuit(3).X(0).X(1), Shots=50, SymbolsDict=SymbolsDict()
)
experiment = ObservableExperiment(
    AnsatzCircuit=ansatz_circuit,
    ObservableTracker=ObservableTracker(qubit_pauli_operator),
)
\end{lstlisting}

In the following we discuss the definition of a \texttt{MitEx} object in
several cases. In each we will use the \texttt{ObservableExperiment} defined
above. In all cases an experiment returns a \texttt{QubitPauliOperator} object
containing an expectation value for each \texttt{QubitPauliString}. 

\subsubsection{Default \texttt{MitEx} Example}
\label{sec:code mitex default}

In its default version, a \texttt{MitEx} object will append a measurement
circuit for each \texttt{QubitPauliString} to the ansatz circuit and execute it
through the pytket \texttt{Backend} the \texttt{MitEx} object is defined by. In
particular the example code snippet:
\begin{lstlisting}[language=Python]
from qermit import MitEx
from pytket.extensions.qiskit import IBMQEmulatorBackend

device_backend = IBMQEmulatorBackend('ibm_lagos', hub='', group='', project='')
mitex = MitEx(backend=device_backend)
mitex_results = mitex.run([experiment])
print(mitex_results[0])

graph = mitex.get_task_graph()
\end{lstlisting}
produces the output:
\begin{lstlisting}
{(Zq[1], Zq[2]): -0.920000000000000}
\end{lstlisting}
with \texttt{graph} being the \texttt{TaskGraph} of \cref{fig:qermit mitex
taskgraph}. Note that as we have used a noisy simulator, the returned value is
different from the ideal value of $-1$.

\subsubsection{ZNE \texttt{MitEx} in Qermit}
\label{sec:code mitex zne}

The ZNE
\texttt{MitEx} has the required inputs: \texttt{backend}, the backend on which
the circuits will be run; 
and \texttt{noise\_scaling\_list}, a list of
factors by which the noise is scaled.  Optionally the type of folding used
can be specified via the
\texttt{\_folding\_type} keyword argument, which expects a \texttt{Folding}
object. The fit used to extrapolate results can be specified via the
\texttt{\_fit\_type} keyword argument, which expects a \texttt{Fit} object.

The example code snippet:
\begin{lstlisting}[language=Python]
from qermit.zero_noise_extrapolation import gen_ZNE_MitEx, Folding, Fit

zne_mitex = gen_ZNE_MitEx(
    backend=device_backend,
    noise_scaling_list=[3, 5],
    folding_type=Folding.circuit,
    fit_type=Fit.exponential,
)

mitex_results = zne_mitex.run([experiment])
print(mitex_results[0])

graph = zne_mitex.get_task_graph()
\end{lstlisting}
produces the output:
\begin{lstlisting}
{(Znode[1], Znode[2]): -0.986666667477977}
\end{lstlisting}
with \texttt{graph} being
the \texttt{TaskGraph} of \cref{fig:qermit zne taskgraph}. Note that ZNE had
returned a value closer to the ideal value of $-1$ than did the example in
\cref{sec:code mitex default} where no error mitigation was used.

\subsubsection{CDR \texttt{MitEx} in Qermit}
\label{sec:code mitex cdr}

The CDR \texttt{MitEx} has required inputs: \texttt{device\_backend}, the
possibly noisy backend used to approximate the expectation of the inputted
circuits; \texttt{simulator\_backend}, an ideal classical simulator;
\texttt{n\_non\_cliffords}, the number of non-Clifford gates in the training
circuits; \texttt{n\_pairs}; and
\texttt{total\_state\_circuits}, the total number of training circuits to use.

The example code snippet:
\begin{lstlisting}[language=Python]
from qermit.clifford_noise_characterisation import gen_CDR_MitEx

cdr_mitex = gen_CDR_MitEx(
    device_backend=device_backend,
    simulator_backend=AerBackend(),
    n_non_cliffords=2,
    n_pairs=2,
    total_state_circuits=50,
)

mitex_results = cdr_mitex.run([experiment])
print(mitex_results[0])

graph = cdr_mitex.get_task_graph()
\end{lstlisting}
produces the output:
\begin{lstlisting}
{(Zq[1], Zq[2]): -1.00000000000000}
\end{lstlisting}
with \texttt{graph} being
the \texttt{TaskGraph} of \cref{fig:qermit cdr taskgraph}. Here we again see an
improvement on the result from \cref{sec:code mitex default}.

\section{Backend Properties}
\label{sec:emulator}

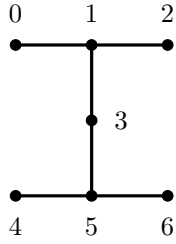
\begin{figure}
    \centering
    \begin{tikzpicture}
        \draw[very thick] (0,0) -- (2,0);
        \filldraw[black] (0,0) circle (2pt) node[below=5]{$4$};
        \filldraw[black] (1,0) circle (2pt) node[below=5]{$5$};
        \filldraw[black] (2,0) circle (2pt) node[below=5]{$6$};

        \draw[very thick] (1,0) -- (1,2);
        \filldraw[black] (1,1) circle (2pt) node[right=5]{$3$};

        \draw[very thick] (0,2) -- (2,2);
        \filldraw[black] (0,2) circle (2pt) node[above=5]{$0$};
        \filldraw[black] (1,2) circle (2pt) node[above=5]{$1$};
        \filldraw[black] (2,2) circle (2pt) node[above=5]{$2$};
    \end{tikzpicture}
    \caption{\textbf{\texttt{ibm\_lagos} and \texttt{ibmq\_casablanca} coupling
    map.} Vertices in the coupling map correspond to qubits, while edges
    indicate that 2-qubit gates can be acted between the qubits which
    correspond to the vertices connected by the edge.}
    \label{fig:emulator coupling}
\end{figure}

The coupling map, describing the connectivity between the qubits, of the
\texttt{ibm\_lagos} and \texttt{ibmq\_casablanca} devices is as seen in
\cref{fig:emulator coupling}. The gates available on the both devices are:
\begin{equation}
    \label{equ:emulator gates}
    \left\{ \CX{}, \RZ{}, \SX{}, \X{}, \I{} \right\}
\end{equation}
Emulation of the devices requires compilation to the
coupling map described in \cref{fig:emulator coupling}, and a rebasing of the
gates used by the circuits to the limited but universal set of
\cref{equ:emulator gates}. 

Average noise properties reported by the devices over the time our experiments
were conducted are seen in \cref{tab:backend properties lagos,tab:backend
properties casablanca}

\begin{table}
    \begin{tabularx}{\hsize}{XX}
        \toprule
        Property & Average Value \\
        \midrule
        $T_1$ time & $128.1$ \\
        $T_2$ time & $99.50$ \\
        Single qubit error rate & $2.307 \times 10^{-4}$ \\
        \CX{} error rate & $8.973 \times 10^{-3}$\\
        Readout error rate & $1.164 \times 10^{-2}$ \\
        \bottomrule
    \end{tabularx}
    \caption{\textbf{\texttt{ibm\_lagos} device properties.}}
    \label{tab:backend properties lagos}
\end{table}

\begin{table}
    \begin{tabularx}{\hsize}{XX}
        \toprule
        Property & Average Value \\
        \midrule
        $T_1$ time ($\mu$s) & $108.5$\\
        $T_2$ time ($\mu$s) & $124.1$\\
        Single qubit error rate & $1.713 \times 10^{-4}$\\
        \CX{} error rate & $8.075 \times 10^{-3}$ \\
        Readout error rate & $2.081 \times 10^{-2}$ \\
        \bottomrule
    \end{tabularx}
    \caption{\textbf{\texttt{ibm\_casablanca} device properties.}}
    \label{tab:backend properties casablanca}
\end{table}

\section{Additional Results Details}
\label{sec:additional results}

Here we present some additional details about the experiments discussed in
\cref{sec:results}. In particular in \cref{tab:additional results pauli} and
\cref{tab:additional results random} we give the average absolute errors and
absolute error of mitigation, as defined in \cref{def:errors}, for the results
presented in \cref{fig:pauli-hardware} and \cref{fig:random-hardware}
respectively. Note that we present mean values here, rather than medians
as in \cref{fig:pauli-hardware} and \cref{fig:random-hardware}.

\newcolumntype{b}{X}
\newcolumntype{m}{>{\hsize=.6\hsize}X}
\newcolumntype{s}{>{\hsize=.3\hsize}X}

\begin{table*}
    \begin{tabularx}{\hsize}{mm|sss|sss|sss|sss}
        \toprule
              &       & \multicolumn{6}{c}{\texttt{ibm\_lagos}} & \multicolumn{6}{c}{\texttt{ibmq\_casablanca}} \\
              &       & \multicolumn{3}{c}{mirrored} & \multicolumn{3}{c}{un-mirrored} & \multicolumn{3}{c}{mirrored} & \multicolumn{3}{c}{un-mirrored}  \\
        Width & Depth & none  & ZNE   & CDR   & none  & ZNE   & CDR   & none   & ZNE   & CDR   & none  & ZNE   & CDR \\
        \midrule
        2     & 2     & 0.053 & 0.046 & 0.016 & 0.025 & 0.426 & 0.119 & 0.062  & 0.052 & 0.012 & 0.032 & 0.016 & 0.178 \\
        2     & 3     &       &       &       & 0.026 & 0.034 & 0.020 &        &       &       & 0.035 & 0.010 & 0.031 \\
        2     & 4     & 0.078 & 0.104 & 0.005 & 0.070 & 0.110 & 0.098 & 0.079  & 0.033 & 0.052 & 0.068 & 0.029 & 0.047 \\
        2     & 5     &       &       &       & 0.047 & 0.049 & 0.023 &        &       &       & 0.055 & 0.049 & 0.039 \\
        3     & 2     & 0.189 & 0.321 & 0.070 & 0.073 & 0.509 & 0.018 & 0.163  & 0.604 & 0.084 & 0.063 & 0.050 & 0.008 \\
        3     & 3     &       &       &       & 0.227 & 1.152 & 0.051 &        &       &       & 0.237 & 0.384 & 0.232 \\
        3     & 4     & 0.562 & 0.721 & 0.059 & 0.254 & 1.017 & 0.043 & 0.350  & 0.286 & 0.163 & 0.211 & 0.403 & 0.030 \\
        4     & 2     & 0.508 & 0.642 & 0.051 & 0.253 & 0.898 & 0.256 & 0.307  & 0.305 & 0.024 & 0.252 & 0.435 & 0.180 \\
        4     & 3     &       &       &       & 0.495 & 1.367 & 0.180 &        &       &       & 0.359 & 1.222 & 0.455 \\
        5     & 2     & 0.562 & 0.595 & 0.017 & 0.365 & 1.209 & 0.282 & 0.335  & 0.604 & 0.015 & 0.335 & 0.427 & 0.141 \\
        \midrule
        Average &     & 0.325 & 0.404 & 0.036 & 0.184 & 0.677 & 0.109 & 0.216  & 0.314 & 0.058 & 0.165 & 0.302 & 0.134 \\
        \bottomrule
    \end{tabularx}
    \caption{\textbf{Mean absolute error and absolute error of mitigation for \pauli{} experiments presented in \cref{fig:pauli-hardware}.}}
    \label{tab:additional results pauli}
\end{table*}

\begin{table*}
    \begin{tabularx}{\hsize}{mm|sss|sss|sss|sss}
        \toprule
              &       & \multicolumn{6}{c}{\texttt{ibm\_lagos}} & \multicolumn{6}{c}{\texttt{ibmq\_casablanca}} \\
              &       & \multicolumn{3}{c}{mirrored} & \multicolumn{3}{c}{un-mirrored} & \multicolumn{3}{c}{mirrored} & \multicolumn{3}{c}{un-mirrored}  \\
        Width & Depth & none  & ZNE   & CDR   & none  & ZNE   & CDR   & none   & ZNE   & CDR   & none  & ZNE   & CDR \\
        \midrule
        2       & 2 & 0.076 & 0.009 & 0.019 & 0.048 & 0.048 & 0.036 & 0.089 & 0.011 & 0.018 & 0.062 & 0.026 & 0.031 \\
        2       & 3 &       &       &       & 0.071 & 0.032 & 0.090 &       &       &       & 0.110 & 0.046 & 0.037 \\
        2       & 4 & 0.121 & 0.028 & 0.028 & 0.040 & 0.053 & 0.067 & 0.125 & 0.011 & 0.029 & 0.089 & 0.028 & 0.065 \\
        2       & 5 &       &       &       & 0.063 & 0.069 & 0.095 &       &       &       & 0.112 & 0.052 & 0.067 \\
        3       & 2 & 0.089 & 0.040 & 0.051 & 0.119 & 0.095 & 0.136 & 0.116 & 0.042 & 0.035 & 0.082 & 0.042 & 0.057 \\
        3       & 3 &       &       &       & 0.208 & 0.092 & 0.203 &       &       &       & 0.303 & 0.248 & 0.339 \\
        3       & 4 & 0.328 & 0.404 & 0.213 & 0.183 & 0.321 & 0.233 & 0.252 & 0.122 & 0.093 & 0.112 & 0.211 & 0.110 \\
        4       & 2 & 0.232 & 0.048 & 0.203 & 0.246 & 0.669 & 0.285 & 0.177 & 0.055 & 0.270 & 0.144 & 0.068 & 0.185 \\
        4       & 3 &       &       &       & 0.288 & 1.100 & 0.368 &       &       &       & 0.243 & 0.080 & 0.180 \\
        5       & 2 & 0.244 & 0.050 & 0.244 & 0.223 & 0.584 & 0.259 & 0.227 & 0.102 & 0.193 & 0.224 & 0.416 & 0.352 \\
        \midrule
        Average &   & 0.182 & 0.097 & 0.126 & 0.199 & 0.306 & 0.177 & 0.164 & 0.057 & 0.106 & 0.148 & 0.122 & 0.142 \\
        \bottomrule
    \end{tabularx}
    \caption{\textbf{Mean absolute error and absolute error of mitigation for \random{} experiments presented in
    \cref{fig:random-hardware}.}}
    \label{tab:additional results random}
\end{table*}

 \section{Behaviour of EM Methods Under Noise Models}
 \label{ap:noise-model}
 	\subsection{Global Depolarising Noise Model}
 	Depolarising noise acts isotropically on quantum states, hence its effect on error mitigation schemes is straightforward to analyse. It takes any state $\rho$ acting on $N$ qubits to
 	\begin{equation}
 		\mathcal{D}(\rho) = (1-p) \rho  + p \frac{\mathbb{I}}{2^{N}}
 	\end{equation}
 	where $p$ is the depolarisation factor.
 	It commutes with any other linear map so $\D\circ \U  = \U\circ \D$ for any unitary channel $\U$. Suppose that a unitary circuit $U$ contains $d_1$ single qubit gates and $d_2$ two-qubit gates, with average error $p_1$ and $p_2$ respectively.
 	
 	If we model the error as globally depolarising, then we have that for any \emph{traceless} observable $O$ the noisy expected value $\<O\>_N$ (in the limit of infinite shots) depends on the exact value $\<O\>$ via
 	\begin{equation}
 		\<O\>_{N} = (1-p_1)^{d_1} (1-p_2)^{d_2}\<O\>
 		\label{eqn:depnoise}
 	\end{equation}
 	For this simplified noise model, the relative error of mitigation remains constant and is independent of the observable and the specific circuit structure. It depends only on the error mitigation method and the number of noisy gates and their average errors.
 	
 	\subsubsection{Zero Noise Extrapolation - Polynomial Fit}
    \label{ap:noise model poly zne}
 	We assume that the error rates are increased artificially via unitary folding with noise stretching factors $\alpha_0 = 1$, $\alpha_1 = 2k_1+1$ and $\alpha_2 = 2k_2+1$. Then the mitigated expected value under a polynomial fit via Richardson extrapolation is
 	\begin{align}
 		\<O\>_{EM} &= F_0 \<P\>_{\alpha_0} + F_1  \<P\>_{\alpha_1} +  F_2 \<P\>_{\alpha_2}
 	\end{align}
 	where the coefficients satisfy
 	\begin{equation}
 		F_{i} = \prod_{i\neq j}  \frac{\alpha_j}{\alpha_i - \alpha_j}
 	\end{equation}
 	and $\<O\>_{\alpha_i} = \gamma^{\alpha_i} \<O\>$ where $\gamma:=  (1-p_1)^{d_1} (1-p_2)^{d_2}$.
 	Therefore the relative error mitigation factor is
 	\begin{equation}
 		\epsilon_{ZNE, poly} = \frac{1- F_0 \gamma - F_1 \gamma^{\alpha_1} - F_2\gamma^{\alpha_2}}{1-\gamma}
        \label{equ:zne noise scaling analytic}
 	\end{equation}
 	For example, the minimal number of folds e.g $k_1 = 1$ and $k_2=2$ gives $F_0 = \frac{15}{8}$, $F_1 = -\frac{5}{4}$ and $F_2 = \frac{3}{8}$.

    This analytic expression allows us to explore the relationship between the
    relative error of mitigation and the depolarising factor. \cref{fig:zne
    analytic noise scaling} reveals a sharp increase in expected relative error
    of mitigation as the depolarisation factor grows.
	
    \begin{figure}
        \centering
        \begin{subfigure}[b]{\columnwidth}
            \centering
	        \includegraphics[width=0.8\columnwidth]{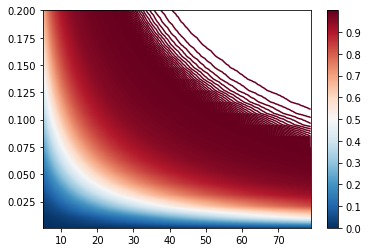}
            \caption{Contour plot of relative error of mitigation
            in terms of $D = d_1 + d_2$ (x-axix) and error rate $p = p_1 = p_2$
            (y-axis) for ZNE with polynomial extrapolation. This is a plot of
            the function seen in \cref{equ:zne noise scaling analytic}.}
            \label{fig:zne analytic noise scaling}
        \end{subfigure}
        \begin{subfigure}[b]{\columnwidth}
            \centering
	        \includegraphics[width=0.8\columnwidth]{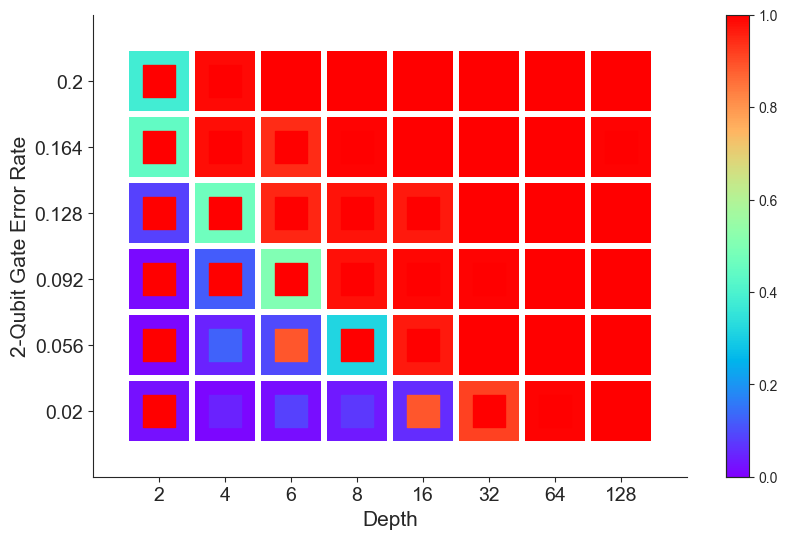}
            \caption{Volumetric plot of relative error of
            mitigation with 4-qubit \pauli{} depth and 2-qubit depolarising factor.
            Exponential fit ZNE is used and 500000 shots are taken per
            circuits. The depolarising factor on 1-qubit gates is a tenth of
            that on 2-qubit gates.}
            \label{fig:zne numeric deep noise scaling}
        \end{subfigure}
        \begin{subfigure}[b]{\columnwidth}
            \centering
	        \includegraphics[width=0.8\columnwidth]{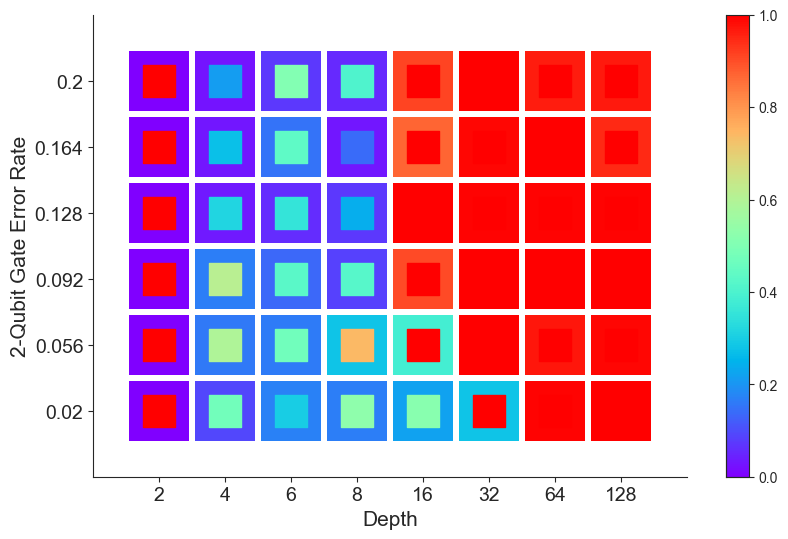}
            \caption{Volumetric plot of relative error of
            mitigation with 4-qubit \pauli{} depth and 2-qubit depolarising factor.
            Exponential fit CDR is used and 500000 shots are taken per
            circuits. The depolarising factor on 1-qubit gates is a tenth of
            that on 2-qubit gates.}
            \label{fig:cdr numeric deep noise scaling}
        \end{subfigure}
        \caption{\textbf{Analytical and numeric calculations of
        relative error of mitigation with changing depolarising factor.}}
        \label{fig:noise scaling}
	\end{figure}
	
 	\subsubsection{Zero Noise Extrapolation -  Exponential Fit}
 	The effect of global depolarising noise model is that for different noise levels $\<O\>_{\alpha_i}  = \gamma^{\alpha_i} \<O\>$. This means that an exponential ZNE fit exactly captures the underlying noise model with the error mitigated observable given by
 	\begin{equation}
 		\<O\>_{EM} = (\<O\>_{\tiny{\alpha_0}}) ^{\frac{\alpha_1}{\alpha_1-\alpha_0}}(\<O\>_{\alpha_1})^{\frac{\alpha_0}{\alpha_0-\alpha_1}}
 	\end{equation}
 	for two levels of noise $\alpha_0, \alpha_1$. Therefore, in the absence of finite sampling errors $\epsilon_{ZNE,exp} = 0$.

 	\subsubsection{Learning with Clifford Data Regression}
 	Clifford-based learning is specifically designed for (global) stochastic Pauli noise; a case in which the linear ansatz exactly matches the noisy expectation values.
 	\begin{equation}
 		\<O\>_{N} = F_1\<O\> + F_0
 	\end{equation}
 	Therefore, in the absence of sampling errors the method should fully correct for the depolarising noise model, and so $\epsilon_{CDR} = 0$.
 
 \subsection{Noisy Classical Simulations}
\label{ap:noisy simulation}

\begin{figure*}
 	\centering

    \includegraphics[width=5cm]{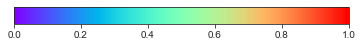}

 	\hfill
 	\begin{subfigure}[b]{0.45\textwidth}
 		\centering
 		\includeplot{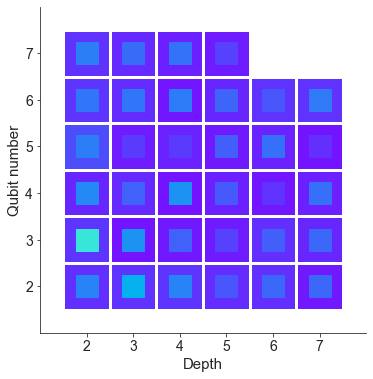}
 		\caption{\random{} mitigated using ZNE.}
 		\label{fig:results depolarising square false zne}
 	\end{subfigure}
 	\hfill
 	\begin{subfigure}[b]{0.45\textwidth}
 		\centering
 		\includeplot{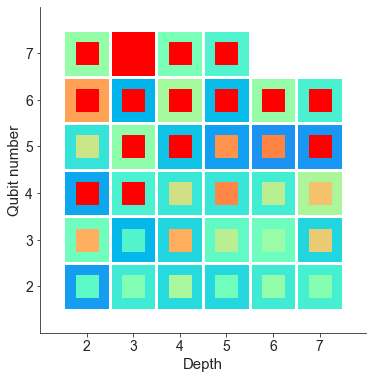}
 		\caption{\random{} mitigated using CDR.}
 		\label{fig:results depolarising square false cdr}
 	\end{subfigure}
 	\hfill
 	
 	\hfill
 	\begin{subfigure}[b]{0.45\textwidth}
 		\centering
 		\includeplot{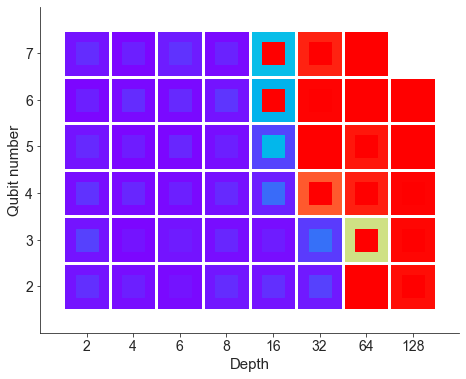}
 		\caption{Mirrored \random{} mitigated using ZNE.}
 		\label{fig:results depolarising square true zne}
 	\end{subfigure}
 	\hfill
 	\begin{subfigure}[b]{0.45\textwidth}
 		\centering
 		\includeplot{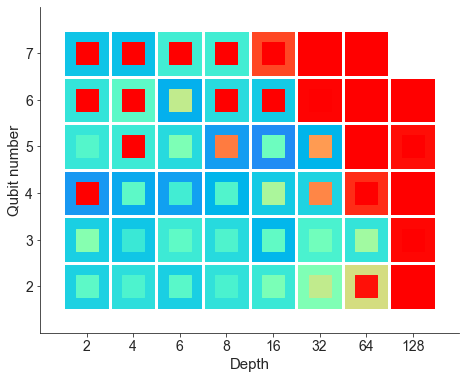}
 		\caption{Mirrored \random{} mitigated using CDR.}
 		\label{fig:results depolarising square true cdr}
 	\end{subfigure}
 	\hfill
 	\caption{\textbf{\random{} run on a depolarising noise simulator}. Each
 		square presents the results of 10 circuits. 500,000 shots are taken in
 		total for each circuit by ZNE and CDR, and 100,000 in the case of the
 		un-mitigated runs.}
 	\label{fig:results depolarising square}
 \end{figure*}

\begin{figure*}
 	\centering

    \includegraphics[width=5cm]{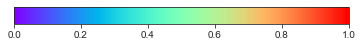}

 	\hfill
 	\begin{subfigure}[b]{0.45\textwidth}
 		\centering
 		\includeplot{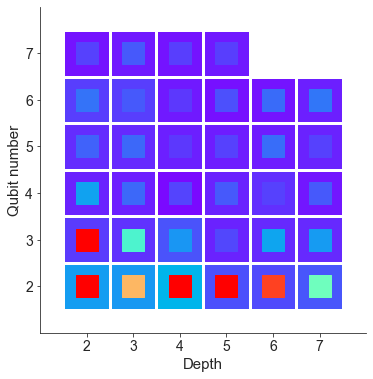}
 		\caption{\pauli{} mitigated with ZNE.}
 		\label{fig:results depolarising deep false zne}
 	\end{subfigure}
 	\hfill
 	\begin{subfigure}[b]{0.45\textwidth}
 		\centering
 		\includeplot{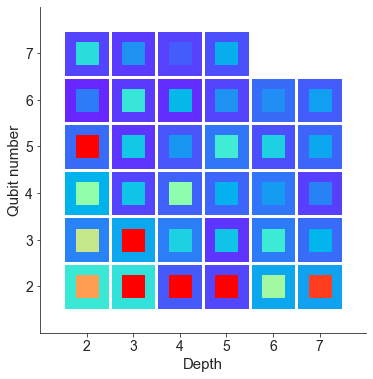}
 		\caption{\pauli{} mitigated with CDR.}
 		\label{fig:results depolarising deep false cdr}
 	\end{subfigure}
 	\hfill
 	
 	\hfill
 	\begin{subfigure}[b]{0.45\textwidth}
 		\centering
 		\includeplot{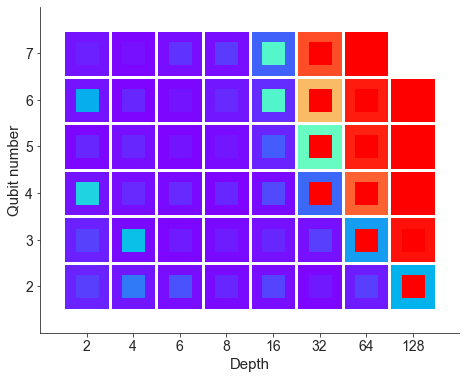}
 		\caption{Mirrored \pauli{} mitigated with ZNE.}
 		\label{fig:results depolarising deep true zne}
 	\end{subfigure}
 	\hfill
 	\begin{subfigure}[b]{0.45\textwidth}
 		\centering
 		\includeplot{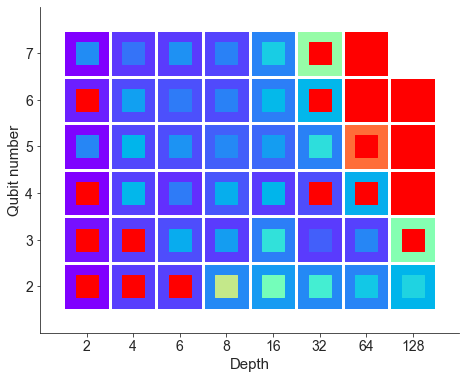}
 		\caption{Mirrored \pauli{} mitigated with CDR.}
 		\label{fig:results depolarising deep true cdr}
 	\end{subfigure}
 	\hfill
 	\caption{\textbf{\pauli{} run on a depolarising noise simulator}. Each
 		square presents the results of 10 circuits. 500,000 shots are taken in
 		total for each circuit by ZNE and CDR, and 100,000 in the case of the
 		un-mitigated runs.}
 	\label{fig:results depolarising deep}
 \end{figure*}

We perform extensive classical simulations of the experiments conducted in
\cref{sec:results}, with the results also discussed there. In particular
depolarising noise simulations with \random{} and \pauli{} are found in
\cref{fig:results depolarising square,fig:results depolarising deep}
respectively.

Besides the fixed depolarising factor simulations presented in
\cref{fig:results depolarising square,fig:results depolarising deep} and
discussed in \cref{sec:results} we additionally explore simulations with
varying noise in \cref{fig:noise scaling}. As discussed in \cref{ap:noise model
poly zne} such an analytical study is possible in the case of ZNE with
polynomial fit, and is presented in \cref{fig:zne analytic noise scaling}. By
comparison, in other areas of this work we consider exponential extrapolation,
which in the idealised case of global depolarising noise results in non zero
relative error of mitigation only as a result of finite sample errors.
\cref{fig:zne numeric deep noise scaling} reveals that for 4 qubit \pauli{} in
the more realistic case of single and two qubit gate local depolarising errors,
and with errors due to sampling, there is a sharp increase in the relative
error of mitigation as the noise level increases. For the same set of
experiments conducted with CDR we again expect 0 relative error of mitigation
in the presence of global depolarising errors and in the absence of finite
sampling errors. As seen in \cref{fig:cdr numeric deep noise scaling} CDR is
less susceptible to finite sampling errors in the presence of local
depolarising noise than is ZNE, speaking to the comparative robustness of the
functional model fitting step of the two schemes.

Finally, emulations additionally classically simulates noise, motivated by
properties of the device. In the case of the emulates experiments shown here
this noise models includes:
\begin{description}
    \item[Readout error:] Single qubit readout errors on measurements.
    \item[Single-qubit errors:] depolarizing error followed by a thermal
        relaxation error on each qubit the gate acts on.
    \item[Two-qubit gate errors:] depolarizing error followed by single-qubit
        thermal relaxation error on each qubit participating in the gate.
\end{description}

\begin{figure*}
 	\centering

    \includegraphics[width=5cm]{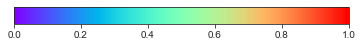}

 	\hfill
 	\begin{subfigure}[b]{0.45\textwidth}
 		\centering
 		\includeplot{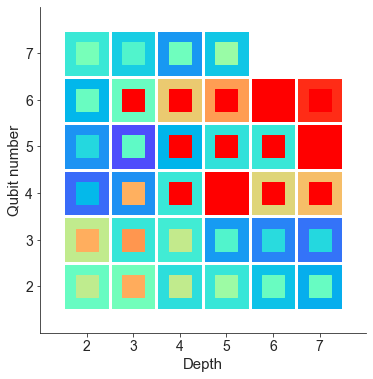}
 		\caption{\random{} mitigated with ZNE.}
 		\label{fig:results emulator square false zne}
 	\end{subfigure}
 	\hfill
 	\begin{subfigure}[b]{0.45\textwidth}
 		\centering
 		\includeplot{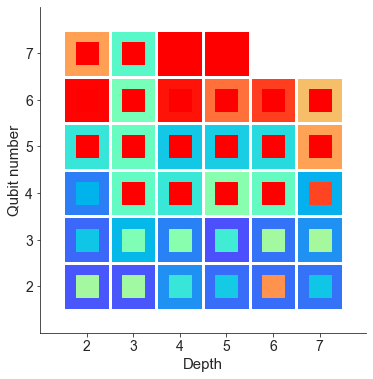}
 		\caption{\random{} mitigated with CDR.}
 		\label{fig:results emulator square false cdr}
 	\end{subfigure}
 	\hfill
 	
 	\hfill
 	\begin{subfigure}[b]{0.45\textwidth}
 		\centering
 		\includeplot{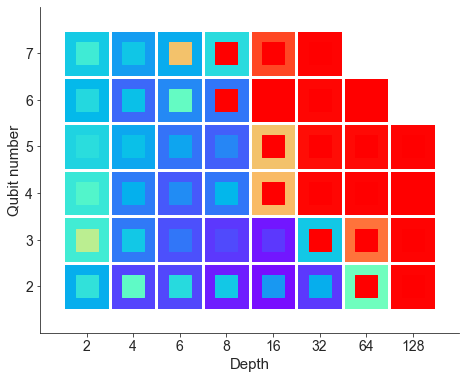}
 		\caption{Mirrored \random{} mitigated with ZNE.}
 		\label{fig:results emulator square true zne}
 	\end{subfigure}
 	\hfill
 	\begin{subfigure}[b]{0.45\textwidth}
 		\centering
 		\includeplot{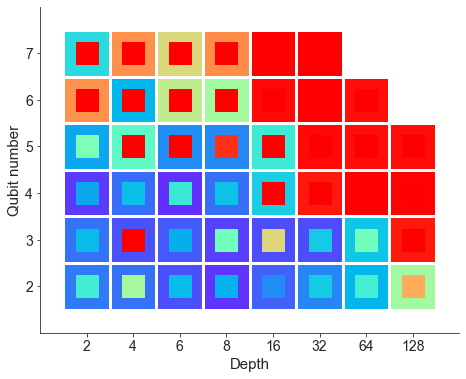}
 		\caption{Mirrored \random{} mitigated with CDR.}
 		\label{fig:results emulator square true cdr}
 	\end{subfigure}
 	\hfill
 	\caption{\textbf{\random{} on emulator backend.} Each square presents the results of
 		10 circuits. 160,000 shots are taken in total for each circuit by ZNE and
 		CDR, and 32,000 in the case of the un-mitigated runs.}
    \label{fig:results emulator square}
 \end{figure*}

\begin{figure*}
 	\centering

    \includegraphics[width=5cm]{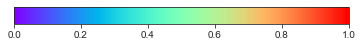}

 	\hfill
 	\begin{subfigure}[b]{0.45\textwidth}
 		\centering
 		\includeplot{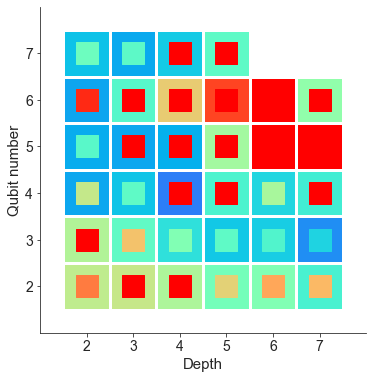}
 		\caption{\pauli{} mitigated with ZNE.}
 		\label{fig:results emulator deep false zne}
 	\end{subfigure}
 	\hfill
 	\begin{subfigure}[b]{0.45\textwidth}
 		\centering
 		\includeplot{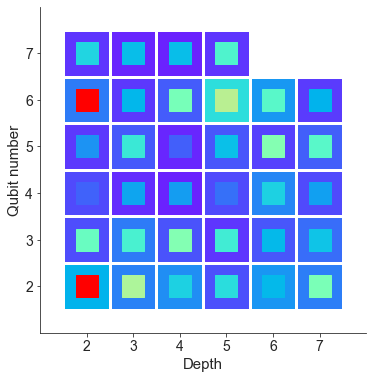}
 		\caption{\pauli{} mitigated with CDR.}
 		\label{fig:results emulator deep false cdr}
 	\end{subfigure}
 	\hfill
 	
 	\hfill
 	\begin{subfigure}[b]{0.45\textwidth}
 		\centering
 		\includeplot{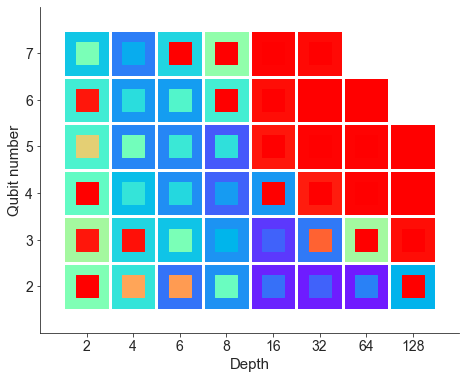}
 		\caption{\pauli{} mitigated with ZNE.}
 		\label{fig:results emulator deep true zne}
 	\end{subfigure}
 	\hfill
 	\begin{subfigure}[b]{0.45\textwidth}
 		\centering
 		\includeplot{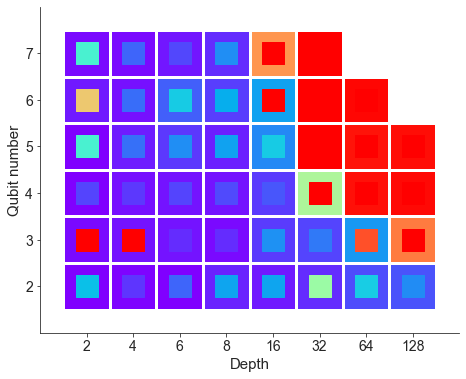}
 		\caption{\pauli{} mitigated with CDR.}
 		\label{fig:results emulator deep true cdr}
 	\end{subfigure}
 	\hfill
 	\caption{\textbf{\pauli{} on emulator backend.} Each square presents the results of
 		10 circuits. 160,000 shots are taken in total for each circuit by ZNE and
 		CDR, and 32,000 in the case of the un-mitigated runs.}
    \label{fig:results emulator deep}
 \end{figure*}

In particular, emulations with \random{} and \pauli{} are found in
\cref{fig:results emulator square} and \cref{fig:results emulator deep}
respectively.
 
\section{Performance of Error Mitigation }
\subsection{Certifying Error Mitigation}
\label{app:certify}
In practical applications we don't have a priori access to the exact value $\<O\>$, as this is the quantity we aim to estimate in the first place.  Typically proof of principle demonstrations of error mitigation strategies require a classical simulation to compare against, and the measures introduced in \cref{def:errors} also assume that $\<O\>$ can be directly computed.  Therefore, the problem of assessing the performance of error mitigation on a specific circuit/observable in regimes beyond classical simulations becomes similar to that of certifying quantum computations. For the purpose of benchmarking the techniques for increasing depth and qubit number one may employ scalable benchmarks consisting, for example, of mirrored circuits.
{Mirrored circuits} -- with the general form $C(w,d) := U_1 U_1\hc U_2 U_2\hc ... U_d U_d\hc$ where $C_i$ are sampled from a specified class of primitive circuits. Then for any observable the ideal expectation of the target state $\rho_{\rm{mirror}} = C(w,d)\rho_0 C(w,d)\hc$ is $\Tr(O\rho_{\rm{mirror}}) = \Tr(O\rho_0)$. For example, if $O$ is a Pauli observable and $\rho_0$ corresponds to the pure product state $|0\>^{\otimes w}$, then the mirrored circuits will give ideal expected values of either $\{0, 1,-1\}$, which can be efficiently determined. Then the relative error of mitigation for the mirrored circuit 
	\begin{equation}
		\epsilon_{\rm{mirror}}(P) =\frac{|\<P\>_{EM,\tilde{\rho}_{\rm{mirror}}} - \<P\>_{\rho_0}|}{|\<P\>_{N,\tilde{\rho}_{\rm{mirror}}} - \<P\>_{\rho_0}|}
	\end{equation}
	can be used as a proxy for the performance of error mitigation on the target state $\rho = U_1...U_d \,  \rho_0  \, U_1\hc.... U_d\hc$ and observable $P$.

\subsection{Relative Error of Mitigation Under Finite Sampling Statistics}
\label{ap:relative-error}
In \cite{diaz2013existence} the authors derive conditions for when the ratio of two independent normally distributed random variables can itself be approximated by a normal distribution. The following is a direct corollary of that result.
\begin{lemma}
	Let $X, Y$  be two independent normal variables with positive means $\mu_X$, $\mu_Y$ and variances $\sigma_X^{2}$ and $\sigma_Y^2$. Suppose the coefficient of variation $\delta_Y := \frac{\sigma_Y}{\mu_Y}  \leq \lambda< 1 $. $\forall \epsilon>0$ there exists  a value $\eta(\epsilon)$ such that if  $0<\delta_Y\leq \eta(\epsilon)$ then the probability distribution function $F_{Z}$ of $Z= X/Y$ approximates a normal distribution $G$ with mean $\mu = \frac{\mu_X}{\mu_Y}$ and variance 
	\begin{equation}
			\sigma =\frac{\mu_X^2}{\mu_Y^2} \left(\frac{\sigma_X^2}{\mu_X^2} + \frac{\sigma_Y^2}{\mu_Y^2}\right)
	\end{equation}
	with $|F_Z(z) - G(z)|\leq \epsilon$ on the interval $z\in [\mu - \frac{\sigma}{\lambda}, \mu + \frac{\sigma}{\lambda}]$ .
	\label{lemma:ratio}
\end{lemma}

In our case, $X = \<\hat{O}\>_{EM} -\<O\>$ corresponds to the difference between the error mitigated estimator and the exact value and $Y =\<\hat{O}\>_N -\<O\>$ corresponds to the difference between the noisy estimator and exact value of the target expectation value. The expected values are then $\mu_{X} = \mathbb{E} \<O\>_{EM} -\<O\> $ respectively $\mu_Y = \mathbb{E} \<O\>_{N} - \<O\> $ with variances $ Var[\<\hat{O}\>_{EM}] = \sigma_{EM}^2$ and $ Var[\<\hat{O}\>_{N}] = \sigma_{N}^2$. 

Since we are interested in relative error of mitigation and this corresponds to absolute values of $Z$, then one can assume that $\mu_X$ and $\mu_Y$ are positive quantities because if that happens not to be the case for a fixed observable then we have the freedom to redefine the random variable $X$ and $Y$ so as to ensure the positivity constraint. 

In the following section we discuss the remaining condition that controls the signal to noise ratio and how the tension between low levels of noise and high variance in the noisy estimators make it difficult to quantify the performance of error mitigation in this regime.

\subsection{Detrimental Effect of Low Noise Level} 
Therefore the only condition left to ensure that \cref{lemma:ratio} can be directly applied to $\epsilon (O)$ is that the coefficient of variation $\delta_Y = \frac{\sigma_N}{\mathbb{E}\<O\>_N - \<O\> }$ is sufficiently small. 
As in the main text if $\rho = U_1 \,...\, U_d \rho_0 U_d\hc\,...\,U_1\hc$ is the target quantum state $\<O\> = Tr(O \rho)$ and its noisy implementation $\tilde{\rho} := \mathcal{E}_1 \circ \mathcal{U}_1 \circ...\circ \mathcal{E}_d\circ \mathcal{U}_d (\rho_0)$  then  $\mathbb{E}\<O\>_N = \Tr(O \tilde{\rho} )$. Now we are looking at the difference,
\begin{align}
	\mu_N &=|\Tr(O (\rho-\tilde{\rho})| \leq ||O||_{\infty}  ||\rho-\rho_0||_{1} \\ 
	&\leq ||O||_{\infty} || \U_1\circ...\circ \U_d - \E_1\circ \U_1... \circ \E_d\circ \U_d||_{\diamond}
\end{align}
where we have used the Holder inequality in the first bound, and then taken a supremum over all possible $\rho$ to bound by the diamond norm $||\Phi||_{\diamond} = sup_{X: ||X||_1\leq 1} ||(\Phi\otimes \mathbb{I}) (X)||_1 $.
For conditions of \cref{lemma:ratio} to be met we want that  $\mu_N\geq \frac{\sigma_N}{\lambda}$ for some $\lambda\le 1 $. This along with sub-multiplicatively of the diamond norm implies that  
\begin{equation}
	\sigma_N \leq \lambda ||O||_{\infty} d e_{\rm{max}}
	\label{eqn:bound}
\end{equation}
where $e_{\rm{max}}$ is  maximal (gate) error rate.
Since $\sigma_N^{2}$ is the variance of the sample mean estimator for the noisy expectation value $\<\hat{O}\>_{N}$, it will depend on the number shots $n_0$ so that $\sigma_{N} = \frac{\sigma_0}{\sqrt{n_0}}$, where $\sigma_0$ is the variance of an individual  (i.i.d) sample of a single measurement. The condition \cref{eqn:bound} becomes 
\begin{equation}
	\frac{\sigma_0}{\sqrt{n_0}} \leq \lambda \,  d \,  e_{\rm{max}} \, ||O||_{\infty} multiplicatively
\end{equation}
In particular this shows that, if one is to ensure that the relative error of mitigation follows a normal distribution under finite sampling then one must have the above trade-offs between the number of shots and depth. The condition also puts a lower bound on the depth of the circuit.

\subsection{Effect of Finite Sampling on the Classical Optimisation to
Determine Fit Parameters}
\label{ap:finite-sampling-fit}

In this section we consider the effect of finite size sampling on fitting the
parameters of the functional model $\mathfrak{F}$. This type of analysis
depends not only on the form of the functional $\mathfrak{F}$ itself but also
on the classical optimization algorithm used to fit the function with noisy
data $\mathbf{D}$. 

For simplicity, we will assume that the data processing step involves a linear
least square optimisation. Recall the following analytical expression
for the fit parameter in a linear model $y_i= Ax_i +B$ for $K$ data samples
\begin{align}
	\hat{A} &= \frac{ \sum_{i=1}^{K}(x_i - \bar{x}) (y_i-\bar{y})}{\sum_{i=1}^{K} (x_i-\bar{x})^2}\\
	\hat{B} &= \bar{y}- \hat{A}\bar{x}
\end{align}
where the estimators minimise the mean square error $\sum_{i=1}^{K} (y_i -
Ax_i- B)^2$. 

As it proves valuable in calculating $\sigma^2_{EM}$, the variance
in $\<\hat{O}\>_{EM}$, note the following derivation of the variance in the
estimator $\hat{A}$
\begin{align}
    \label{equ:regression slope variance}
    \text{Var} \left[ \hat{A} \right] 
        &= \text{Var} \left[ \frac{ \sum_{i=1}^{K}(x_i - \bar{x}) (y_i-\bar{y})}{\sum_{i=1}^{K} (x_i-\bar{x})^2} \right] \\
        &= \frac{ \sum_{i=1}^{K}(x_i - \bar{x})^2 \text{Var} \left[ (y_i-\bar{y}) \right] }{\left( \sum_{i=1}^{K} (x_i-\bar{x})^2 \right)^2} \\
        &= \frac{ 1 }{\left( \sum_{i=1}^{K} (x_i-\bar{x})^2 \right)} \frac{K+1}{K} \sigma^2
\end{align}
where we have: assumed $\text{Var} \left[ y_i \right] = \sigma^2$ for all $i$,
employed $\bar{y} = \frac{1}{K} \sum_{i=1}^{K}y_i$, and used that $x_i$ and
$\bar{x}$ are constants and not random variables. Note further
\begin{equation}
    \label{equ:regression intercept variance}
    \text{Var} \left[ \hat{B} \right]
        = \text{Var} \left[ \bar{y} - \hat{A} \bar{x} \right]
        = \frac{\sigma^2}{K} + \bar{x}^2 \text{Var} \left[ \hat{A} \right].
\end{equation}

In the examples considered below, for CDR and ZNE the finite size
sampling errors imply that the $y_i$'s can be viewed as normally distributed
random variables.

\subsubsection{Clifford Data Regression}
\label{app:performance parameters cdr}

We consider a linear functional model as in \cref{eqn: CDRmodel} with least
square optimisation as a classical method to determine the model parameters.
Then we have $\hat{A} = \hat{F}_1$, $\hat{B}=\hat{F}_0$, $x_i = D^{c}_{i}$, and
$y_i = D^{q}_i$ and the error mitigated estimator is given by 
\begin{equation}
	\<\hat{O}\>_{EM}  =  \frac{\<\hat{O}\>_N - \hat{F}_0}{\hat{F}_1}.
\end{equation}

If $\hat{F}_0$ and $\hat{F}_1$ were a constant fixed values, $F_0$ and $F_1$,
then 
\begin{equation}
    \label{equ:cdr variance approx}
	\sigma_{EM}^2 = \frac{\sigma_N^2}{F_1^2}.
\end{equation} 
However, both are unbiased estimators constructed out of (noisy)
sample mean estimators, so incur a variance due to a finite number of shots.
Treating $\hat{F}_0$, $\hat{F}_1$ and $\<\hat{O}\>_N$ as normally-distributed independent
random variables gives
\begin{align}
    \sigma_{\text{EM}}^2 
        &= \text{Var} \left[ \frac{\<\hat{O}\>_N -\hat{F}_0}{\hat{F}_1} \right] \\
        &\approx \frac{\sigma_N^2 + \text{Var} \left[ \hat{F_0} \right]}{\hat{F}_1^2} + \frac{\left( \< \hat{O} \>_N - \hat{F}_0\right)^2 \text{Var} \left[ \hat{F_1} \right]}{\hat{F}_1^4} \\
        &= \frac{\left( |D^c| + 1 \right) \sigma_N^2}{|D^c| \hat{F}_1^2} + \frac{\left( \< \hat{O} \>_N - \hat{F}_0\right)^2 + \hat{F}_1^2 \bar{D^c}^2}{\hat{F}_1^4} \text{Var} \left[ \hat{F}_1 \right] \\
        &= \frac{\left( |D^c| + 1 \right) \sigma_N^2}{|D^c| \hat{F}_1^2} \left[ 1 + \frac{ \left( \< \hat{O} \>_N - \hat{F}_0\right)^2 + \hat{F}_1^2 \bar{D^c}^2}{\hat{F}_1^2 \left( \sum_{i=1}^{|D^c|} \left( D^c_i - \bar{D^c} \right)^2 \right)} \right]
\end{align}
where we employ respectively \cref{lemma:ratio}, \cref{equ:regression intercept
variance}, and \cref{equ:regression slope variance}, and use $|D^c|$ to denote
the number of Clifford training circuits. Note that an honest treatment of
\cref{lemma:ratio} would require knowledge of exact values of $\< O \>_N$,
$F_0$, and $F_1$, rather than the estimators we use here. As we do not have
access to these exact values, we conjecture that this approximation is accurate
in the limit of large numbers of shots.

\subsubsection{Zero Noise Extrapolation}
\label{app:performance parameters zne}

We consider exponential extrapolation and the parameters in the functional
model $D_{\lambda_i} = F_0 e^{F_1\lambda_i}$ with $\<\hat{O}\>_{EM} =
\hat{F_0}$ determined via linear least square optimisation. Linearisation of
the exponential function gives $\hat{A} = \hat{F}_1$, $\hat{B} = \log \left(
\hat{F_0} \right)$, $x_i = \lambda_i$, and $y_i = \log \left( D^q_{\lambda_i}
\right)$. If again we treat $\hat{F}_1$ as being the constant value $F_1$ then
we recover from \cref{equ:regression intercept variance} that
\begin{equation}
    \text{Var} \left[ \log \left( \< \hat{O} \>_{EM} \right) \right] = \frac{1}{|\lambda|} \text{Var} \left[ \log \left( \< \hat{O} \>_{N} \right) \right]
\end{equation}
where $|\lambda|$ is the number of noise scaling values, and where we have
assumed 
\begin{equation}
    \text{Var} \left[ \log \left( D^{q}_{\lambda_i} \right) \right] = \text{Var} \left[ \log \left( \< \hat{O} \>_N \right) \right], \; \forall i.
\end{equation}
Assuming further that $\frac{\sigma_{EM}^2}{\< O \>_{EM}^2},
\frac{\sigma_{N}^2}{\< O \>_{N}^2} \ll 1$, gives
\begin{equation}
    \label{equ:zne variance approx}
    \frac{\sigma_{EM}^2}{\< \hat{O} \>_{EM}^2} \approx \frac{1}{|\lambda|} \frac{\sigma_{N}^2}{\< \hat{O} \>_{N}^2}.
\end{equation}
However, it is more honest to consider $\hat{F}_1$ as a random variable,
in which case
\begin{align}
    \frac{\sigma_{EM}^2}{\< \hat{O} \>_{EM}^2}
        &\approx \frac{1}{|\lambda|} \frac{\sigma_{N}^2}{\< \hat{O} \>_{N}^2} + \bar{\lambda} \text{Var} \left[ \hat{F}_1 \right] \\
        &= \frac{1}{|\lambda|} \frac{\sigma_{N}^2}{\< \hat{O} \>_{N}^2} + \bar{\lambda} \frac{ 1 }{\left( \sum_{i=1}^{|\lambda|} (\lambda_i-\bar{\lambda})^2 \right)} \frac{\lambda+1}{|\lambda|} \sigma_{N}^2
\end{align}

\subsubsection{Relative Error of Mitigation Variance}

\begin{figure}
    \centering
    \includegraphics[width=\linewidth]{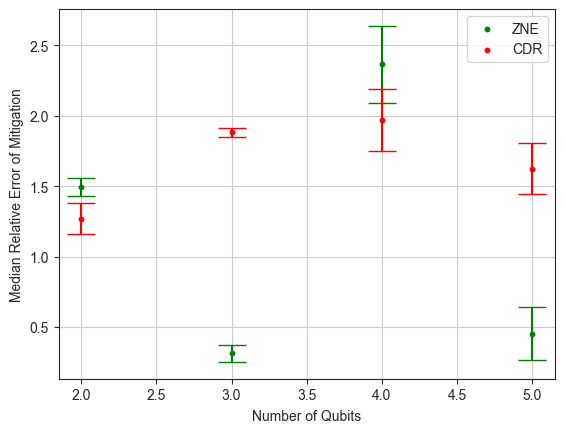}
    \caption{Data is as in \cref{fig:results random lagos false zne} and
    \cref{fig:results random lagos false cdr}, with a qubit number of two.
    Error bars show one standard deviation.}
    \label{fig:relative error variance}
\end{figure}

We combine the arguments in \cref{app:performance parameters
cdr,app:performance parameters zne} with \cref{equ:relative error variance} to
produce \cref{fig:relative error variance}, which exemplifies well controlled
variance in the case of our experiments. Here $\sigma^2_N$, the variance in the
estimator of the noisy expectation value, is approximated by resampling the
shots produced in the experiments of \cref{fig:results random lagos false
zne,fig:results random lagos false cdr}. $\sigma^2_{EM}$ is calculated using
\cref{equ:cdr variance approx,equ:zne variance approx}.

\end{document}